\documentclass[nonatbib]{elsarticle}
\usepackage[utf8]{inputenc}
\usepackage[british]{babel}
\usepackage{csquotes}
\usepackage{subfiles}
\usepackage{amsmath}
\usepackage{eurosym}
\usepackage{booktabs}
\usepackage{graphicx}
\usepackage{multirow}
\usepackage{lscape}
\usepackage{setspace}
\usepackage[table,xcdraw]{xcolor}
\usepackage[hidelinks]{hyperref}
\hypersetup{colorlinks=true,citecolor={blue},citebordercolor={blue}}
\usepackage{geometry}
\makeatletter
\let\c@author\relax
\makeatother
\usepackage[backend=biber, style=numeric-comp, autocite=superscript, maxcitenames=1, isbn=false, date=year, url=false, sorting=none]{biblatex}
\DeclareAutoCiteCommand{superscript}{\supercite}{\supercites}
\AtEveryBibitem{\clearlist{language}}
\makeatletter
\def\ps@pprintTitle{%
  \let\@oddhead\@empty
  \let\@evenhead\@empty
  \let\@oddfoot\@empty
  \let\@evenfoot\@oddfoot}
\makeatother

\title{Effects of system-blind prosumers in energy models}

\author[1,2,3]{Adeline Guéret\corref{cor1}}

\author[1]{Wolf-Peter Schill}

\author[1,2]{Felix Schmidt}

\cortext[cor1]{Corresponding author: \texttt{agueret@diw.de}.}
\affiliation[1]{organization={DIW Berlin},
addressline={Mohrenstraße 58},
city={10117 Berlin},
country={Germany}}

\affiliation[2]{organization={Technische Universität Berlin},
addressline={Straße des 17.\ Juni 135},
city={10623 Berlin},
country={Germany}}

\affiliation[3]{organization={OFCE - Sciences Po Paris},
addressline={10, Place de Catalogne},
city={75014 Paris},
country={France}}

\begin{document}

\begin{keyword}
     Prosumers \sep Battery electric vehicles \sep Energy system modelling \sep Sector coupling \sep Renewable energy
\end{keyword}

\begin{abstract}
Prosumer households that generate and store electricity from rooftop PV installations play an increasing role in electricity markets around the world. As retail tariffs usually do not convey time-varying wholesale price signals to households and the rollout of smart meters is low in many countries, prosumers do not necessarily self-consume and feed-in solar electricity in a system-friendly way. The effects of such system-blind behaviours are typically neglected in energy system models, which rarely account for prosumers. In this paper, we embed a calibrated self-generation constraint into a linear capacity expansion model to approximate the incentives of prosumers to minimise their electricity bills. We apply our method to a German case study for 2030 featuring sector coupling with battery electric vehicles. We show that parametrising the self-generation constraint such that the prosumer electricity bill is as low as possible approximates prosumer decisions well for a broad range of tariff schemes. Based on this, we quantify distortions that might arise in energy models that do not account for prosumers. For our case study, we find that the optimal battery storage capacity increases by up to 200\% if prosumer constraints are included. The main driver is the imperfect substitutability between home and utility-scale batteries. We conclude that energy system models could benefit from implementing this straightforward method. 
\end{abstract}

\maketitle
\renewcommand{\thefootnote}{\alph{footnote}}

\doublespacing

\section{Introduction}

Around the globe, households increasingly install rooftop photovoltaic (PV) systems\autocite{renewables2024analysis}. They are referred to as ``prosumers'', reflecting their two-fold identity as both producers and consumers of electricity. Small home battery storage systems installed in combination with rooftop PV panels, which is sometimes described as ``prosumage''\autocite{schill2017prosumage}\textsuperscript{,}\footnote{In what follows, we use prosumers and prosumage interchangeably to refer to households generating electricity, whether equipped with a home battery or not.}, are also spreading. In recent years, technological developments bringing down prices for PV modules and batteries have made prosumage substantially more attractive in many countries. In Europe, the~2022~energy crisis, which led to sizeable electricity price spikes, has further increased profitability prospects of prosumage. In some countries like Belgium and the Netherlands, prosumers already represent almost 25\% to 30\% of households while for most European countries this share lies between 2 and 5\%\autocite{acer2024retail}. In absolute terms, Germany gathers the most prosumer households in Europe with around 2.5 million households. 

While prosumers already play a noticeable role in some European countries, their importance will likely further grow across the continent in the future. First of all, residential rooftop PV installations are expected to grow due to the sustained decline in PV module prices as well as relatively short installation times\autocite{solar2023europe}. Furthermore, rooftop PV has the potential to also reduce grid expansion costs\autocite{rahdan2024distributed, steinbach2024enabling}. Most importantly, the electrification of households' energy end uses such as space and water heating or mobility via the rollout of heat pumps and battery electric vehicles (BEVs) will offer new opportunities for prosumers to benefit from increased amounts of self-generated electricity that they consume behind-the-meter. Increased price volatility on the electricity market\autocite{schoniger2022comes} might also strengthen households' willingness to hedge themselves against these uncertainties by investing in their own production capacities.

It is likely that prosumers aim to minimise their own electricity bills by deciding how much capacity to install and when to consume electricity\autocite{semmelmann2024empirical}. In doing so, they are guided by retail tariffs and, in many countries, also by feed-in tariffs. If smart-metering technologies, dynamic retail pricing schemes and dynamic feed-in tariffs are not available, which is still the default setting nowadays in most European countries, wholesale price signals have little influence on prosumers' investment and consumption decisions. Prosumers might thus make capacity and consumption decisions that deviate from what central planner (CP) cost minimisation models would predict. Hence, energy system models not integrating prosumers' characteristics may lead to distorted results. In order to avoid such distortions in results, which may affect optimal dispatch and investment outcomes of various technologies, it is necessary that energy system models take into account prosumers that aim to minimise their electricity bills.

However, fully and consistently including prosumer optimisation problems in a central planner optimisation problem remains complex, due to the difficulty to capture the full interaction loop between prosumers and the central electricity system. Formally, this leads to equilibrium problems which are hard to solve numerically. Indeed, prosumers' decisions both are influenced by and influence investment and dispatch decisions of other power market participants. The former effect strongly hinges upon the regulatory framework in place. Previous studies have investigated how electricity prices and various retail pricing schemes impact prosumers' investment and consumption decisions\autocite{hoppmann2014economic, lang2016profitability, kaschub2016solar, schwarz2018self, say2018coming, keiner2019cost, stute2024assessing}. The latter effect derives from the fact that prosumers' behind-the-meter activities impact the magnitude and temporal shape of the overall electricity demand, while the electricity fed back to the grid also modifies the overall electricity supply\autocite{agnew2015effect, schill2017prosumage}. Hence, some studies have investigated the effects of prosumers' decisions on the power sector, employing various methods. Using a scenario-based analysis, \cite{yu2018prospective} investigates how the deployment of residential rooftop PV panels might impact the optimal capacity mix for France in 2030, while \cite{fares2017impacts} look at the impact of operating decentralised home batteries on peak power demand, peak power injections and overall electricity consumption. Other studies implement a sequential approach\autocite{breyer2018solar, say2020degrees}, first determining prosumers' optimal decisions for given tariff structures, and inputting them in a second step to power sector models, to quantify how prosumers might impact optimal PV or storage capacities in the power system. While these approaches offer an improvement over not considering prosumers in energy models, they do not iterate until some convergence is reached between the assumptions passed to the prosumers in the first step and the outcomes of the second stage, e.g., wholesale electricity prices. In an attempt to integrate more closely both perspectives, \cite{gunther2021prosumage} use a mixed complementarity problem to link the optimality conditions of the prosumer electricity bill minimisation problem and the power sector cost minimisation problem to investigate the impact of different tariff schemes on optimal investments and hourly use of PV-battery systems. This approach however remains difficult to solve numerically and is thus limited to a dispatch-only setting of the rest of the power sector. \cite{sarfarazi2023improving} couple an energy system model with an agent-based model and analyse prosumers' impact on the optimal generation mix and dispatch in Germany for different tariff options. While this approach effectively bridges the gap between both perspectives, it involves modelling efforts that are not easily transferable to other models. Another way of interlinking prosumers and the central planner is to include additional constraints in the central planner optimisation problem that approximate the prosumer perspective to some extent. For instance, some works propose to include a self-consumption constraint\autocite{schill2017decentralized, schick2020role} and investigate how power sector costs are affected when including this constraint or not.
 
We build on the latter approach and include a self-generation constraint in a linear capacity expansion model to circumvent the complexity of an equilibrium problem while still taking into account interactions between prosumers and the power sector. In addition to residential rooftop PV panels and home batteries, we also consider battery electric vehicles and their interactions with the other distributed technologies as well as the central electricity system. To the best of our knowledge, such a sector-coupled prosumer framework has not been considered in previous studies integrating prosumers and the central planner so far, despite its relevance for future highly renewable energy systems. For a German case study where prosumers can realise self-consumption benefits, we determine the self-generation rate that minimises the electricity bill of prosumers and carry out checks, in which we compare prosumers' outputs in the integrated system approach with their outputs when minimising their own electricity bills. Ultimately, we investigate to what extent various model outcomes differ when including prosumer constraints or not. 

Our results show that the method employed approximates well prosumer bill minimisation in our sector-coupled German case study. Hence, the method implemented, while interlinking prosumers and the central planner more closely than in sequential approaches, effectively allows to circumvent the formulation of an equilibrium problem. We further find that approximating prosumers leads to higher overall battery storage capacities in the power sector. This is explained by prosumers operating batteries with a view to supply an optimal share of their domestic load rather than providing flexibility to the central electricity system. We conclude that not taking into account prosumers in highly renewable energy system models leads to substantially underestimate the total installed short-duration storage capacity in the system. Hence, insights derived from energy system models would benefit from following this simple method whenever possible.

\section{Methods}
\label{sec:methods}

Under the assumptions of perfect competition, symmetric information, rational agents and no externalities, economic theory predicts that a welfare maximising central planner of a power system reaches the same outcome as a long-run market equilibrium\autocite{mas_colell_microeconomic_1995}. This equivalence permits abstracting from individual market participants and formulating a power system model as an optimisation problem. An implicit assumption is that all actors maximise their utility against the same system price, which they take as given. However, depending on the regulatory framework, the latter assumption does not hold for prosumers which face retail tariffs that differ from system prices. Indeed, across different countries, the design of retail tariffs takes various forms, ranging from time-invariant volumetric tariffs to real-time prices including dynamic network charges. Likewise, many countries remunerate the feed-in of rooftop solar PV at time-invariant tariffs, i.e. independent from wholesale prices in respective hours\autocite{ceer2023res}. 

Under such tariff designs, the equivalence of a central planner problem and a frictionless market outcome breaks down. One option to resolve this would be to model the interaction between the optimisation problems of prosumers and other actors in the system explicitly. This can be done in form of mixed complementarity or bilevel problems, depending on the assumed nature of the interaction\autocite{gabriel2013complementarity}. Such non-linear and non-convex problems are NP-hard and inherently difficult to solve at scale. The fact that the prosumer-system interaction yields products of system dual variables (the system price entering the retail tariff calculation) and prosumer primal variables (the amount bought from the grid) further complicates numerical solution approaches.

The approach presented in this section seeks to avoid an equilibrium formulation by amending the standard optimisation framework, which is used in many models of power sectors and energy systems, by a set of constraints that forces the central planner to approximate prosumers' decisions as if they were made in view of retail and feed-in tariffs and in disregard of wider system effects.

\subsection{Central planner optimisation problem}

To evaluate the impact of not taking prosumers into account in the central planner problem, we use the open-source capacity expansion model DIETER\autocite{zerrahn2017long-run,gaete2021dieterpy}. It is a simple linear model that features a standard objective function which minimises overall investment and dispatch costs of the power sector under various feasibility and operational constraints. Different versions of this model have been used to investigate the role of electricity storage\autocite{schill2018long-run}, electric heating\autocite{roth2024power}, battery electric vehicles\autocite{gaetemorales2024power,gueret2024impacts}, green hydrogen\autocite{kirchem2023power}, as well as prosumers\autocite{schill2017prosumage,say2020degrees} in renewable energy transition scenarios. More details about model assumptions and the parametrisation used here are provided in Section~\ref{sec:model_parameters}.

\begin{figure}[!ht]
    \centering
    \includegraphics[width=0.85\linewidth]{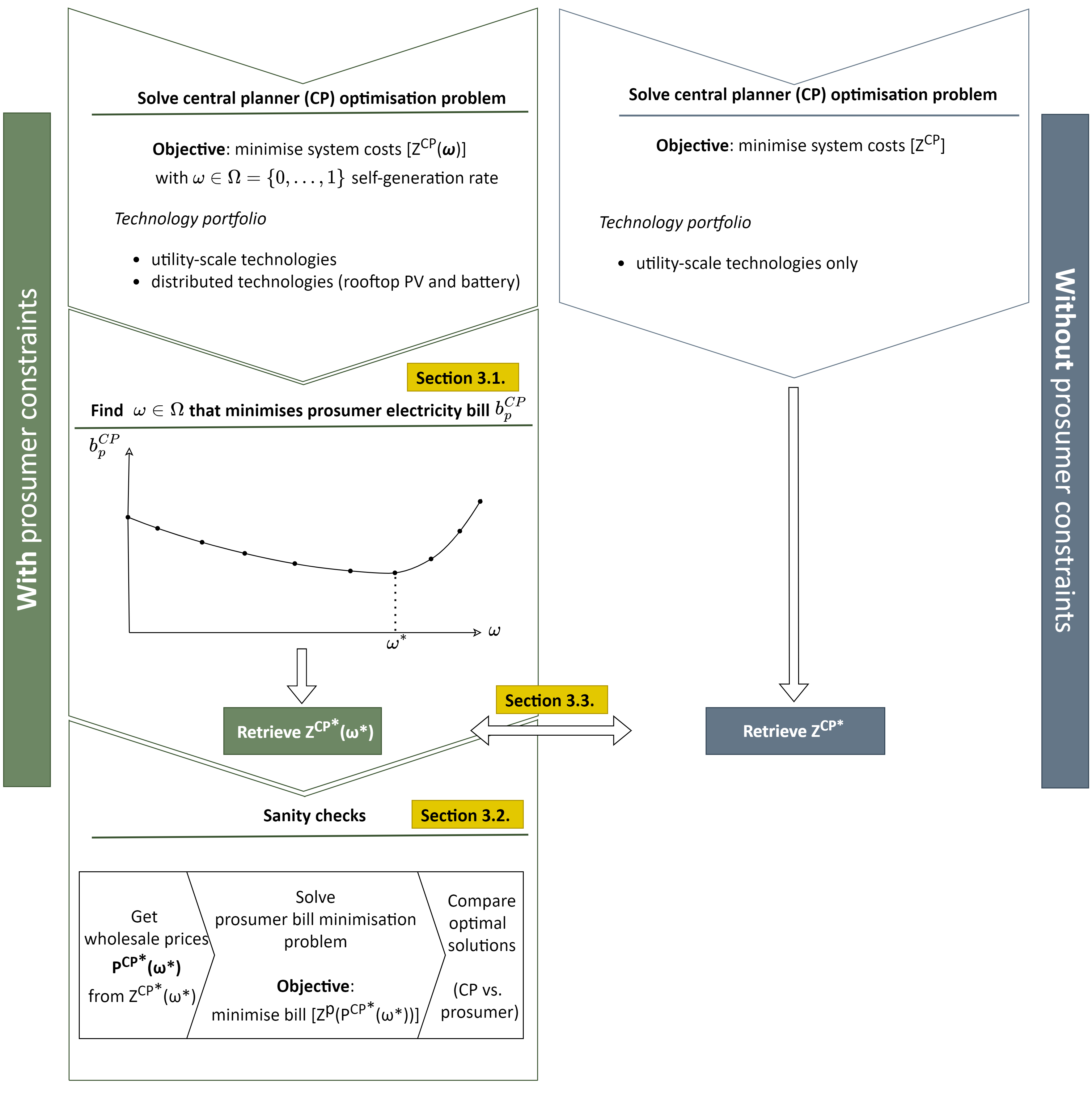}
    \caption{Workflow overview}
    \label{fig:steps_overview}
\end{figure}

Here, we amend the central planner optimisation problem with a component that approximates prosumers which minimise their electricity bills as detailed below. We then compare optimal model solutions when including or not including the prosumer approximation. An overview of the workflow is depicted in Figure~\ref{fig:steps_overview}. 

\begin{figure}[!ht]
    \centering
    \includegraphics[width=0.45\linewidth]{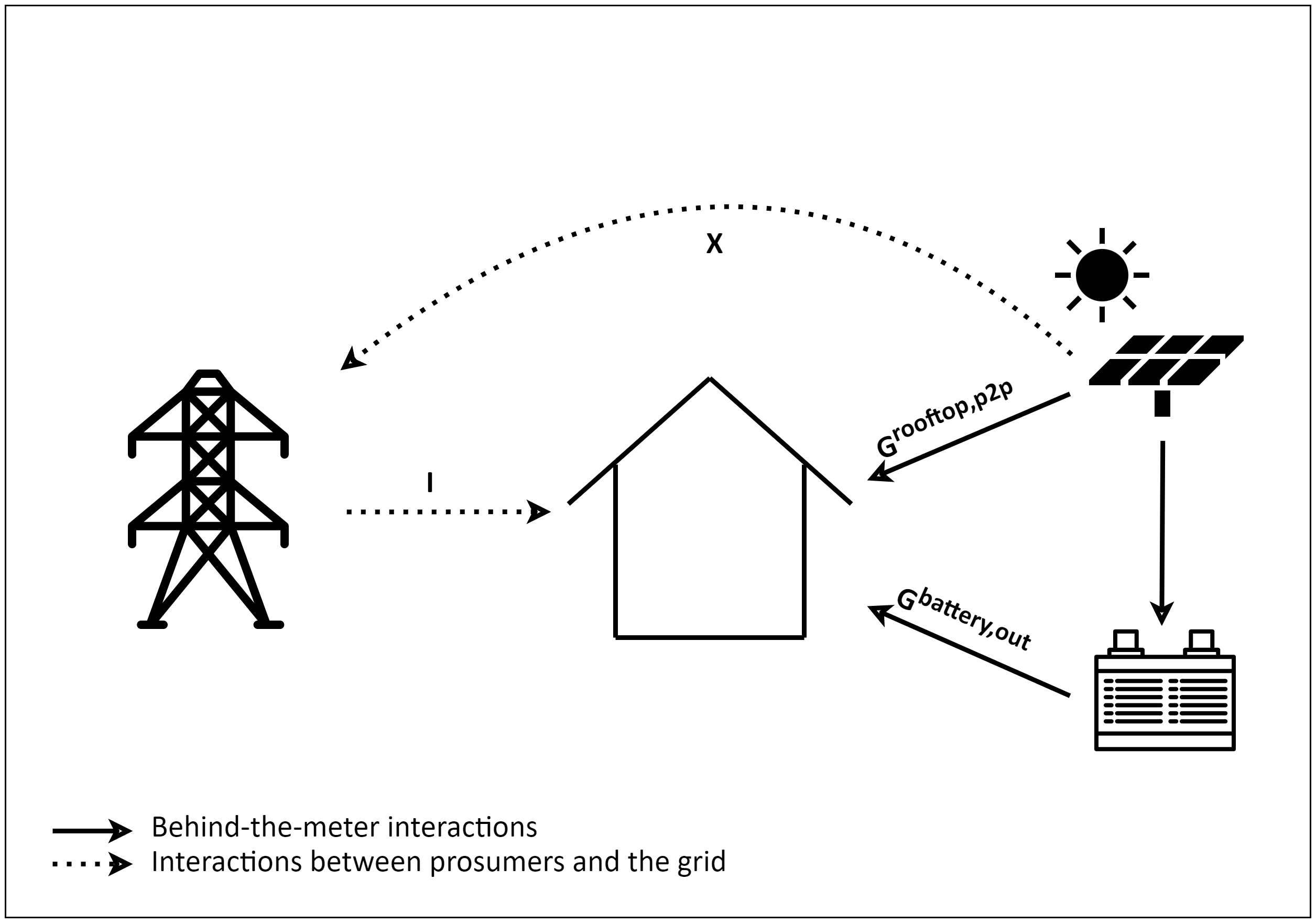}
    \includegraphics[width=0.45\linewidth]{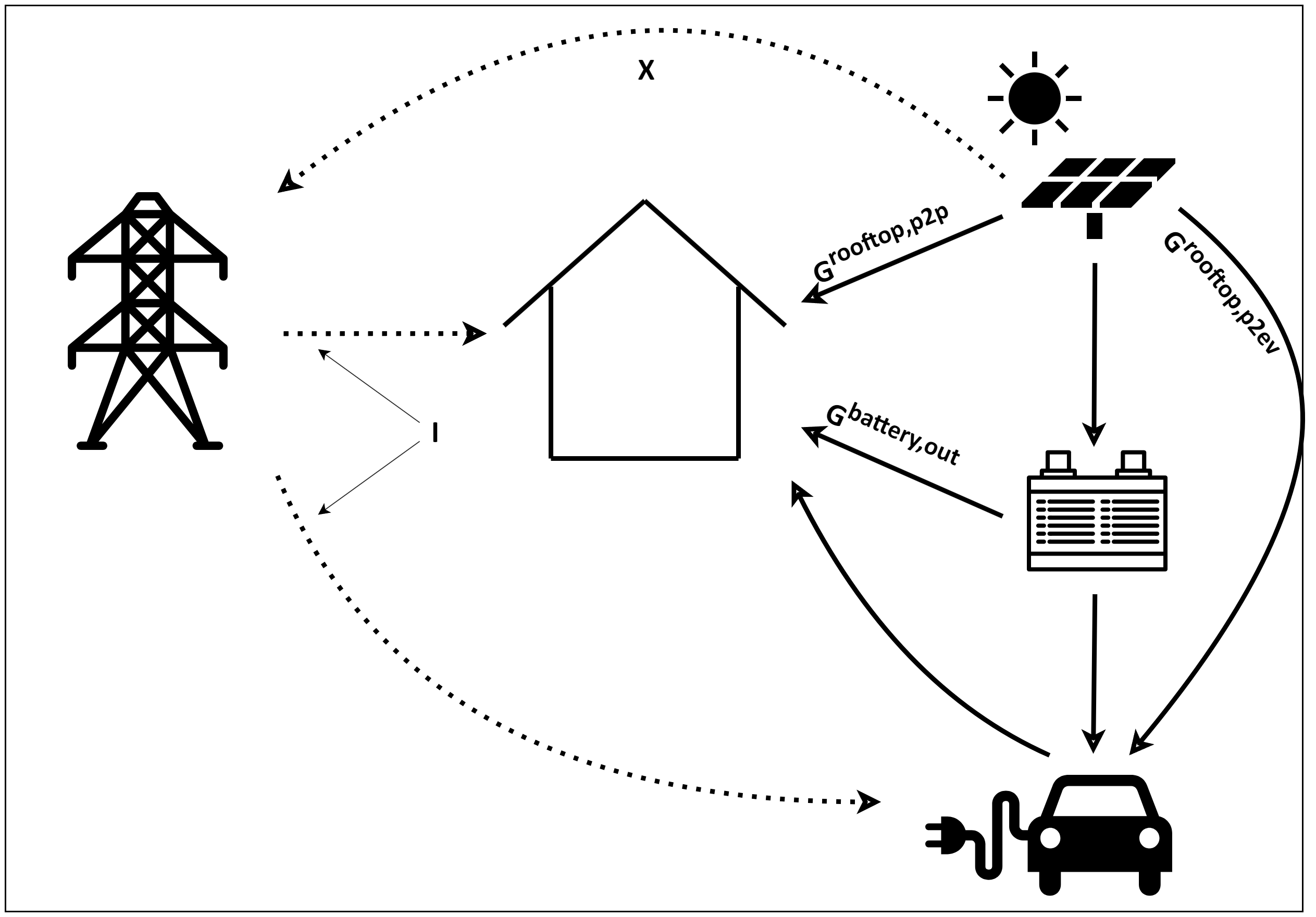}
    \caption{Electricity flows in the two model setups \textsc{No BEVs} (left) and \textsc{With BEVs} (right)}
    \label{fig:flowchart}
\end{figure}

In the model without prosumer constraints, only utility-scale technologies are considered. When approximating the prosumer bill minimisation, we introduce additional distributed rooftop PV and home battery storage systems. The prosumer load can then be supplied either by self-generated electricity or by electricity drawn from the grid. Prosumers can consume self-generated electricity ``on the spot'' as it is generated by rooftop PV panels or later by charging and discharging the home battery. Self-generated electricity that is neither directly consumed nor stored by prosumers is fed into the grid. We assume that the home battery storage cannot interact with the grid (see Figure~\ref{fig:flowchart}). This assumptions reflects the current situation of most PV-battery systems in Germany.

We further differentiate two model setups, \textsc{No BEVs} and \textsc{With BEVs}. In \textsc{With BEVs}, prosumers have a battery electric vehicle in addition to their rooftop PV and home battery installations. BEVs can charge from the grid whenever they are parked at home or parked away and connected to a charging station. When parked at home, self-generated electricity can also be used to charge the BEV. The home battery can also discharge self-generated electricity to charge the BEV. Finally, the BEV can discharge to supply the prosumer load, also referred to as vehicle-to-home (V2H). For simplicity, we do not allow BEVs to discharge to the grid nor to the home battery storage. 

\subsection{Self-generation constraint}

The central planner problem neither considers electricity retail tariffs that have to be paid by prosumers when drawing electricity from the grid nor feed-in tariffs that are paid to prosumers when they feed electricity back to the grid. As rooftop PV is more expensive than utility-scale PV and since potential grid benefits of rooftop PV are ignored, the cost-minimal model solution would not contain any distributed rooftop PV or home battery capacities. This is what most energy models typically find, with a few notable exceptions\autocite{keiner2019cost, rahdan2024distributed}. In reality, however, households do have incentives to install rooftop PV panels and home batteries as substituting grid consumption with self-generated electricity as well as selling surplus electricity at feed-in tariffs can help them to lower their electricity bills. In order to approximate prosumers minimising their bills, we constrain the central planner to build distributed capacities that serve the prosumer load behind-the-meter by introducing a self-generation constraint. 

In \textsc{No BEVs}, this constraint, formalised in equation~(\ref{eq:sgc_no_bevs}), enforces that the aggregate annual amount of self-generated power from rooftop PV panels that is not fed into the grid should be at least as high as a share $\omega$ of the prosumer total annual load,~$d^{p}$. In what follows, we refer to $\omega$ as the ``self-generation rate''.  

\begin{equation}
    \sum_{h \in \mathcal{H}} \left( G^{rooftop,p2p} _{h} +  G^{battery,out} _{h} \right) \geq \omega \times \sum_{h \in \mathcal{H}} d^{p} _{h} \quad \text{with}~\mathcal{H} = [\![ 1,8760 ]\!]
\label{eq:sgc_no_bevs}
\end{equation}

Here, $ G^{rooftop,p2p} _{h}$ represents electricity generated from rooftop PV directly supplying prosumer load at hour~$h$ and $G^{battery,out} _{h}$ the home battery power outflow supplying prosumer load, net of discharging losses. 

In \textsc{With BEVs}, the self-generation constraint is modified on both sides, as shown in equation~(\ref{eq:sgc_with_bevs}). On the right-hand side, the annual electricity consumption for driving the electric vehicle,~$d^{p,ev}$ is added to the prosumer domestic load. On the left-hand side, we consider the overall amount of self-generated electricity that supplies the prosumer domestic load or flows into the battery of the BEV ($G_{h} ^{rooftop,p2ev}$). We subtract self-generated electricity that is lost in charging and discharging processes of both the home battery storage and the BEV, $L$, as it does not meaningfully contribute to self-generation. Likewise, electricity fed into the grid does not enter the left-hand side of the self-generation constraint.

\begin{equation}
    \sum_{h \in \mathcal{H}} \left( G^{rooftop,p2p} _{h} +  G^{battery,out} _{h} + G^{rooftop,p2ev} _{h} - L_{h} \right) \geq \omega \times \sum_{h \in \mathcal{H}} \left( d^{p} _{h} + d^{p,ev} _{h} \right) \quad \text{with}~\mathcal{H} = [\![ 1,8760 ]\!]
\label{eq:sgc_with_bevs}
\end{equation}

\subsection{Prosumer total electricity bill}

While the self-generation constraint enforces that the central planner optimisation problem leads to capacity investment in rooftop PV panels, it can lead to vastly different optimal solutions depending on how $\omega$ is parametrised. Hence, we perform a grid search on the parameter space~$\Omega = \{0,\dots,1\}$ of the self-generation rate~$\omega$. We then compute the prosumer electricity bill for each value of~$\omega$ and select the value~$\omega^*$ that minimises the bill. In doing so, we aim to find a solution that reconciles prosumer objectives and overall system cost minimisation.  

\begin{equation}
    \begin{aligned}
       b^{CP} _{p} (\omega) &= \sum_{d \in \mathcal{D}} \text{capex}^{d} (\omega) + \sum_{d \in \mathcal{D}} \sum_{h \in \mathcal{H}} \text{opex}^{d} _{h} (\omega) + \sum_{h \in \mathcal{H}} \left( r_{h} (\omega) i_{h}  (\omega) - f x_{h}  (\omega) \right)  \\
                      &\text{with}~\mathcal{D} = \{\text{rooftop}, \text{battery}\}~\text{and}~\mathcal{H} = [\![ 1,8760 ]\!]
    \end{aligned}
\label{eq:bill_syspro}
\end{equation}

The prosumer electricity bill $b^{CP} _{p} (\omega)$, shown in equation~(\ref{eq:bill_syspro}), comprises overall capacity expenditures ($\text{capex}$) for building rooftop PV panels and home batteries as well as costs for operating them ($\text{opex}$), the latter including both fixed and variable costs. Besides, the prosumer electricity bill also includes net purchase costs, i.e. overall costs paid by prosumers when drawing electricity from the grid ($i$, as ``imports'') net of the revenues they generate when feeding excess self-generated electricity back to the grid ($x$, as ``exports''), assuming a time-invariant feed-in tariff $f$. Retail electricity prices $r_{h}$ depend on wholesale prices that are also given by the central planner optimal solution and assumptions on different types of tariff adders. More details about retail pricing schemes are provided in Section~\ref{sec:scenarios} and in the Supplemental Notes. Lastly, operational costs also include BEV battery degradation costs related to V2H operations in the setup \textsc{With BEVs}. Note that all these costs are determined by the central planner optimal solution, $Z^{CP^{*}}(\omega)$.

\subsection{Sanity checks}

To evaluate whether the previous steps effectively allow to approximate prosumer behaviour, we perform sanity checks which consist in comparing optimal outcomes for prosumer variables in the central planner problem to the optimal outcomes of an isolated prosumer bill minimisation problem. The objective function of the latter as given in equation~(\ref{eq:bill_pro}) is to minimise the electricity bill defined as the sum of capacity and operational expenditures and net purchase costs. Retail prices $r_h$ depend on electricity wholesale prices $P^{CP^{*}}(\omega^{*})$, which are exogenously given by the central planner optimal solution $Z^{CP^{*}} (\omega^*)$ as in the previous step. However, investment $N$ and dispatch~$G$ for prosumer technologies are now decision variables, as well as grid consumption (or imports) $I$ and grid feed-in (or exports) $X$. 

\begin{equation}
    \begin{aligned}
        \min_{N,G,I,X} B^p (P^{CP^{*}}(\omega^*)) &= \sum_{d \in \mathcal{D}} \text{capex}(N^d) + \sum_{d \in \mathcal{D}} \sum_{h \in \mathcal{H}} \text{opex}^{d} _{h}(N^d,G^d) + \sum_{h \in \mathcal{H}} \left( r_{h} (P^{CP^{*}}(\omega^*)) I _{h} - f X_{h} \right)  \\
                         &\text{with}~\mathcal{D} = \{\text{rooftop}, \text{battery}\}~\text{and}~\mathcal{H} = [\![ 1,8760 ]\!]
    \end{aligned}
\label{eq:bill_pro}
\end{equation}

The outcomes from the prosumer bill minimisation problem differ from the central planner optimisation problem which approximates the prosumer bill minimisation since the latter does not consider the price frictions created by retail and feed-in tariffs. While the central planner problem implicitly assumes that prosumers operate in a way that contributes to minimising overall system costs, optimal prosumer decisions in the electricity bill minimisation problem are guided by retail and feed-in tariffs. This can lead to diverging optimal results in terms of capacity investments and dispatch for prosumer technologies, electricity bills and self-generation rates.  

\subsection{Model parametrisation}
\label{sec:model_parameters}

We calibrate the model to a 2030 scenario for Germany. Electricity generation technologies considered are closed or open cycle gas turbines (CCGT and OCGT), hard coal, biomass, utility-scale PV, run-of-river hydro, wind onshore and wind offshore. Electricity storage technologies considered are stationary lithium-ion batteries, closed or open pumped-hydro storage as well as hydro reservoirs. We also include the option of a long-duration electricity storage technology which comprises converting electricity to green hydrogen via electrolysis, which is then compressed and stored in underground caverns before being reconverted to electricity using open cycle gas turbines. Rooftop PV panels are assumed to be around twice as expensive as utility-scale PV\autocite{danish_energy_agency_technology_2025}. We further assume that an average prosumer household cannot install more than 15~kWp of rooftop PV. Cost and technology parameters are detailed in Tables~\ref{tab:parameters_gen}~and~\ref{tab:parameters_sto}. 

Hourly capacity factor time series for solar, wind onshore and offshore as well as inflow time series for run-of-river, open pumped hydro and reservoirs are from the weather year 2017-2018 (July~1 until June~30)\autocite{antonini2024weather}. The model is solved for all consecutive 8760~hours of that period. For simplicity, we focus on Germany only and abstract from interconnections with neighbouring countries. The base electricity demand is considered fully price-inelastic and amounts to around 518~TWh for the full year\autocite{wiese2019open} after correction for electrified heat, which is taken into account separately, using the When2Heat\autocite{ruhnau_update_2022} dataset and after accounting for industry electrification\autocite{neumann_potential_2023}. We assume a total annual load of a prosumer household of 5~MWh and use standard load profiles for household consumers for the year 2015\autocite{stromnetzberlin_netznutzer}. 

Additionally, we include electricity demand related to future sector coupling technologies such as heat pumps, green hydrogen used in other sectors, and electric vehicles. The yearly demand for electrified heat supplied by heat pumps amounts to~291 TWh, based on \cite{m_jrc-idees-2021_2024} and \cite{antonini2024weather}. The model also includes a yearly demand for green hydrogen from other sectors of 42~TWh which is assumed to be evenly distributed over all hours and has to be satisfied by hydrogen produced via domestic electrolysis. For BEVs, we use the open source probabilistic tool \textit{emobpy}\autocite{gaete2021open} to generate time series of electricity demand and charging availability. Underlying parameters used to calibrate the tool are detailed in the Supplemental Notes. Battery electric vehicles are assumed to be charging flexibly, i.e., whenever it incurs the lowest system costs, as long as batteries are sufficiently charged for the next car trip. More particularly, we assume that they can charge not only when parked at home but also when parked away from home. To avoid model artifacts, we assume that the price for charging away corresponds to the retail tariff that prosumers face. Besides, we assume that they cannot discharge to the grid but can, whenever parked at home, discharge to meet the prosumer load.

\subsection{Scenarios}
\label{sec:scenarios}

For each setup, we compare optimal solutions for the central planner problem with or without prosumer constraints. We consider different numbers of prosumer households (2,~5~or~10~million) and various retail electricity pricing schemes. 

Reference cases are problems without prosumer constraints, yet they differ slightly depending on the setup considered. In \textsc{No BEVs}, there is only one reference. In \textsc{With BEVs}, there are three respective references as these include 2,~5 or 10~million BEVs. This amounts to a yearly driving electricity consumption of around 5~TWh, 14~TWh or 27~TWh, respectively. In the corresponding problems that approximates prosumers, only prosumers are assumed to have BEVs.

Electricity fed into the grid generates a revenue for prosumers as they receive a volumetric feed-in compensation, which is assumed to be time-invariant. Electricity drawn from the grid by prosumers is valued at retail tariffs which consist of two components. One component reflects the costs of electricity procurement and supply, which depends on $\omega$ and stems from the optimal solution of the central planner problem. The other component is a tariff adder that reflects other costs such as taxes, network charges, metering and other surcharges and is assumed to be independent from $\omega$. As there is substantial uncertainty about the future level of volumetric tariff adders, we consider different values of~10, 15, 20~and~25~\euro~cts/kWh. The~2024~average non-energy part of Germany's retail electricity price is about~23.5~\euro~cts/kWh\footnote{\url{https://www.bundesnetzagentur.de/DE/Vportal/Energie/PreiseAbschlaege/Tarife-table.html}}). 

Aside from time-invariant (``Fixed'') volumetric retail pricing, we also model two alternative tariff schemes, time-of-use pricing (ToU) and real-time pricing (RTP). For each of these two schemes, we consider three variants: (i)~the first variant assumes that only the procurement part is time-varying, while the tariff adder remains time-invariant; (ii)~the second variant assumes that the procurement part and half of the tariff adder are time-varying, while the other half remains time-invariant; (iii)~the third variant assumes that both the procurement and the tariff adder components are fully time-varying. More details on the assumed tariff designs are provided in the Supplemental Notes. 

\section{Results}
\label{sec:results}

\subsection{The self-generation rate that minimises the prosumer electricity bill} 

\begin{figure}[!ht]
    \centering
    \includegraphics[width=0.9\linewidth]{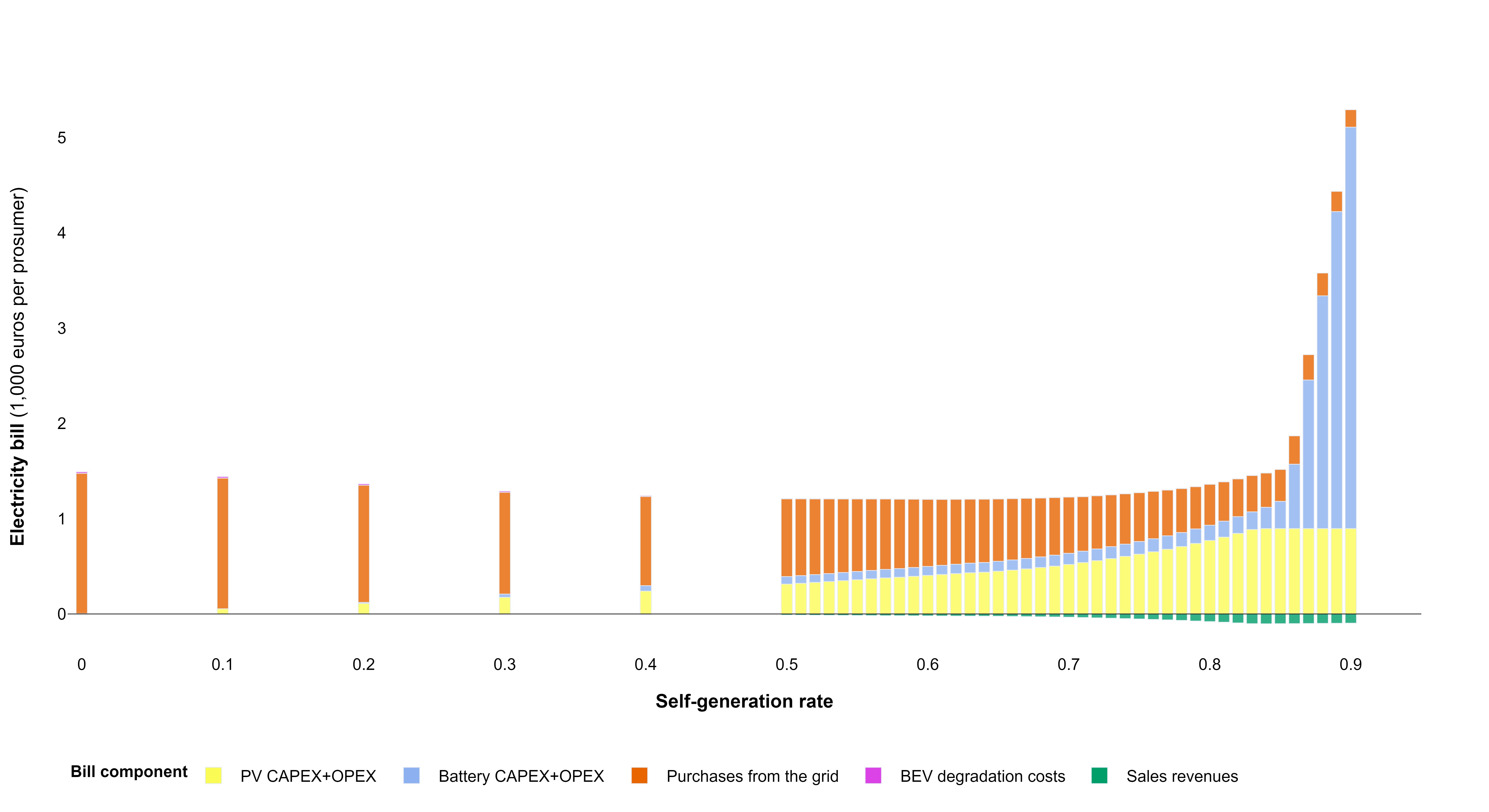}
    \caption{Decomposition of the prosumer annual electricity bill by cost component [\textsc{With BEVs}, time-invariant tariffs, tariff adder of 20 \euro cts/kWh, 10 million prosumers]}
    \label{fig:costs_decomposition_ev}
\end{figure}

For all pricing schemes, tariff adder values and number of prosumers considered, the yearly prosumer electricity bill is a convex function of the self-generation rate $\omega$ (see Figures~\ref{fig:costs_total_all_noev}~and~\ref{fig:costs_total_all_withev}). In particular, the bill does not reach its lowest values at the boundaries of the parameter space. The convexity of the prosumer total costs is driven by the interaction between grid consumption costs (net of feed-in revenues), on the one hand, and rooftop PV capacity expenditures on the other (Figure~\ref{fig:costs_decomposition_ev} for \textsc{With BEVs} and Figure~\ref{fig:costs_decomposition_std} for \textsc{No BEVs}). Up to the cost-minimising~$\omega$, every additional kWh of self-consumed electricity is on average cheaper than a kWh consumed from the grid. Past the cost-minimising~$\omega$, the decrease in grid purchase costs is overcompensated by growing expenditures for additional rooftop PV and home battery capacity. This is because the matching of rooftop PV time profiles and residual prosumer load gets marginally worse with increasing~$\omega$. Above a very high self-generation threshold, the only way of achieving higher self-generation rates is to invest in very large home battery storage capacities, which dramatically drives up costs.

We observe that the higher the tariff adder, the higher the cost-minimising self-generation rate~$\omega$ (see Fi\-gure~\ref{fig:cost_minimising_omega_noev} and Figure~\ref{fig:cost_minimising_omega_withev}). This is because higher tariff adders make it more profitable for prosumers to substitute grid electricity with self-generated electricity, even if facilitating the latter requires larger batteries and thus incurs higher costs. Across all settings except RTP 100, which we treat separately as explained hereafter, optimal self-generation rates lie between~$0.5$ and~$0.8$.

\subsection{Comparing optimal decisions in central planner and prosumer problems} 
\label{sec:checks}

\begin{figure}[!ht]
    \centering
    \includegraphics[width=1.0\linewidth]{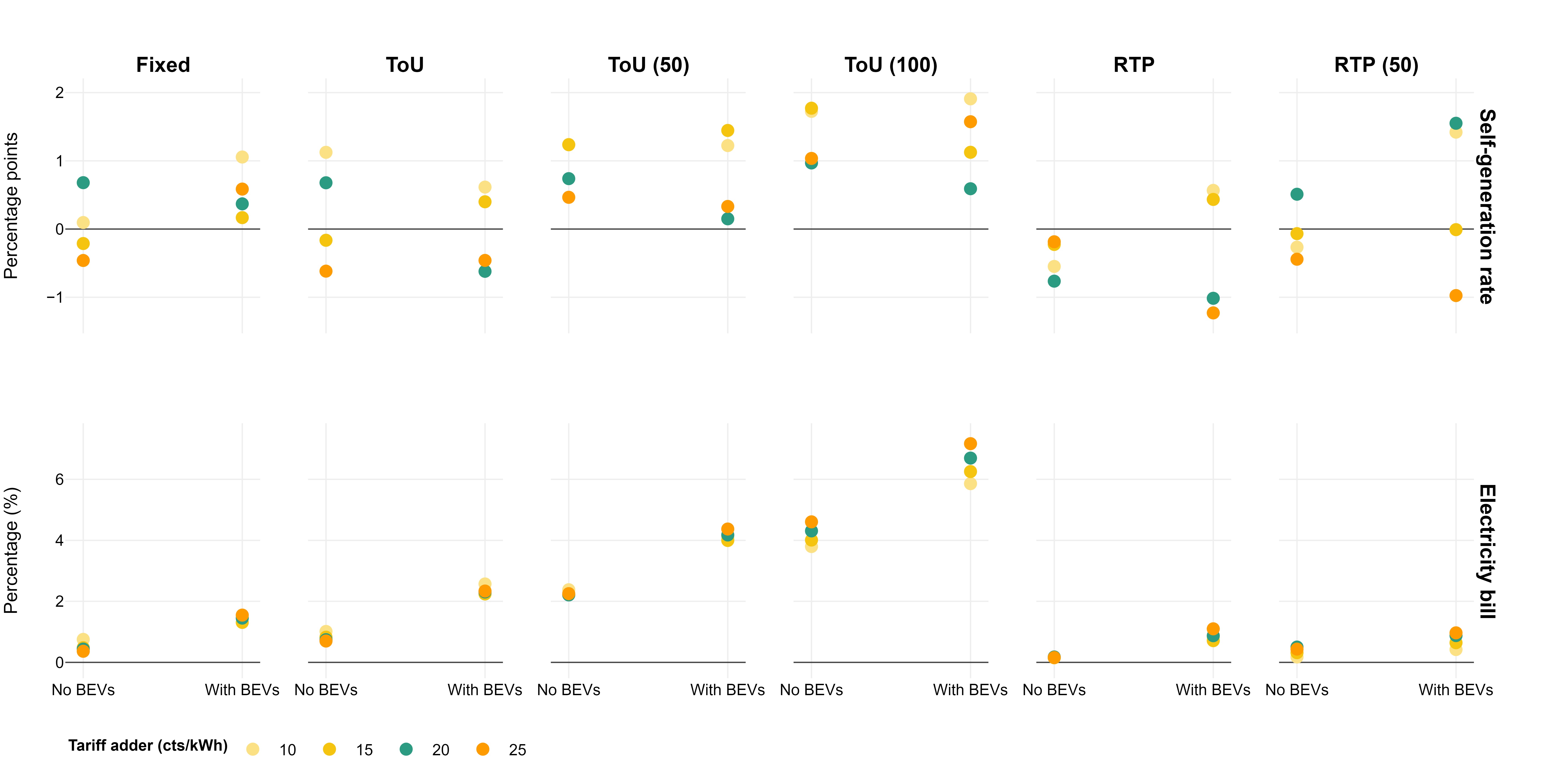}
    \caption{Deviation of the bill-minimising self-generation rate (upper panel) and the prosumer electricity bill (lower panel) between the central planner optimisation problem and the prosumer optimisation problem [10 million prosumers]. Note: a positive value means that the outcome in the central planner problem is larger than the outcome in the prosumer problem. Electricity bill deviations are expressed relative to the optimal outcome in the prosumer bill minimisation problem.}
    \label{fig:diff_sgr_costs_10mio}
\end{figure}

\begin{figure}[!ht]
    \centering
    \includegraphics[width=1.0\linewidth]{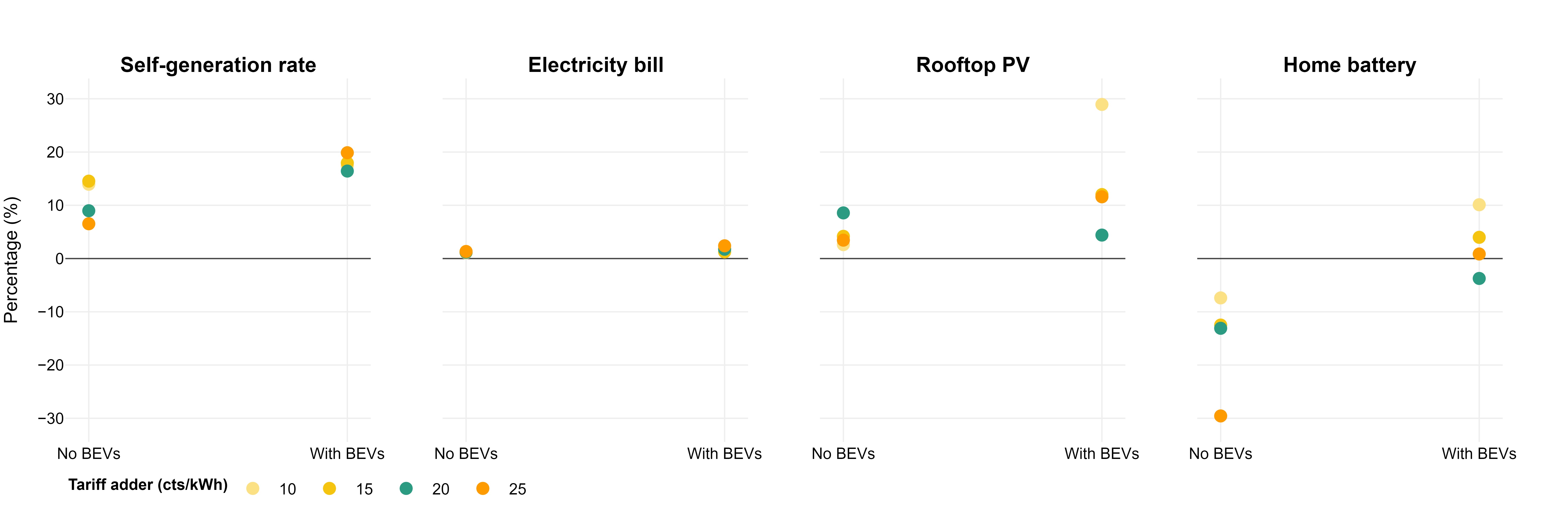}
    \caption{Deviations between the central planner optimisation problem and the prosumer optimisation problem for various variables [RTP 100, 10 million prosumers]. Note: a positive value means that the outcome in the central planner problem is larger than the outcome in the prosumer problem. Deviations are expressed relative to the optimal outcome in the prosumer bill minimisation problem (in percentage), except for the self-generation rate where deviations are expressed in percentage points.}
    \label{fig:diff_rtp100p_10mio}
\end{figure}

The difference between the bill-minimising self-generation rates derived from the central planner problem that approximates prosumers and the isolated prosumer optimisation problem for given tariffs is very small for all pricing schemes except for the fully time-varying variant of real-time pricing (RTP 100). We opt to exclude the latter from the set of tariffs going forward and discuss it separately at the end of this section (see Figure~\ref{fig:diff_rtp100p_10mio}). For the remaining pricing schemes, the optimal self-generation rate of the central planner deviates by less than two percentage points from the self-generation rate derived from the prosumer problem (see Figure~\ref{fig:diff_sgr_costs_10mio}, upper panel). This deviation does not go systematically in the same direction, meaning that the prosumer approximation leads to overestimating the self-generation rate in some cases, and underestimating it in others. 

Differences in annual prosumer electricity bill also remain small, below~five percent, in all but one cases (see Figure~\ref{fig:diff_sgr_costs_10mio}, lower panel). If electric vehicles are included, deviations tend to be more pronounced as charging and discharging of BEV allows for more potential grid interactions of prosumers. Moreover, deviations are always positive. This is expected since any approximation of the bill minimisation problem will lead to sub-optimal decisions for the prosumer. That being said, the small deviations we observe show that the central planner still approximates well the annual prosumer electricity bill.   

\begin{figure}[!ht]
    \centering
    \includegraphics[width=1.0\linewidth]{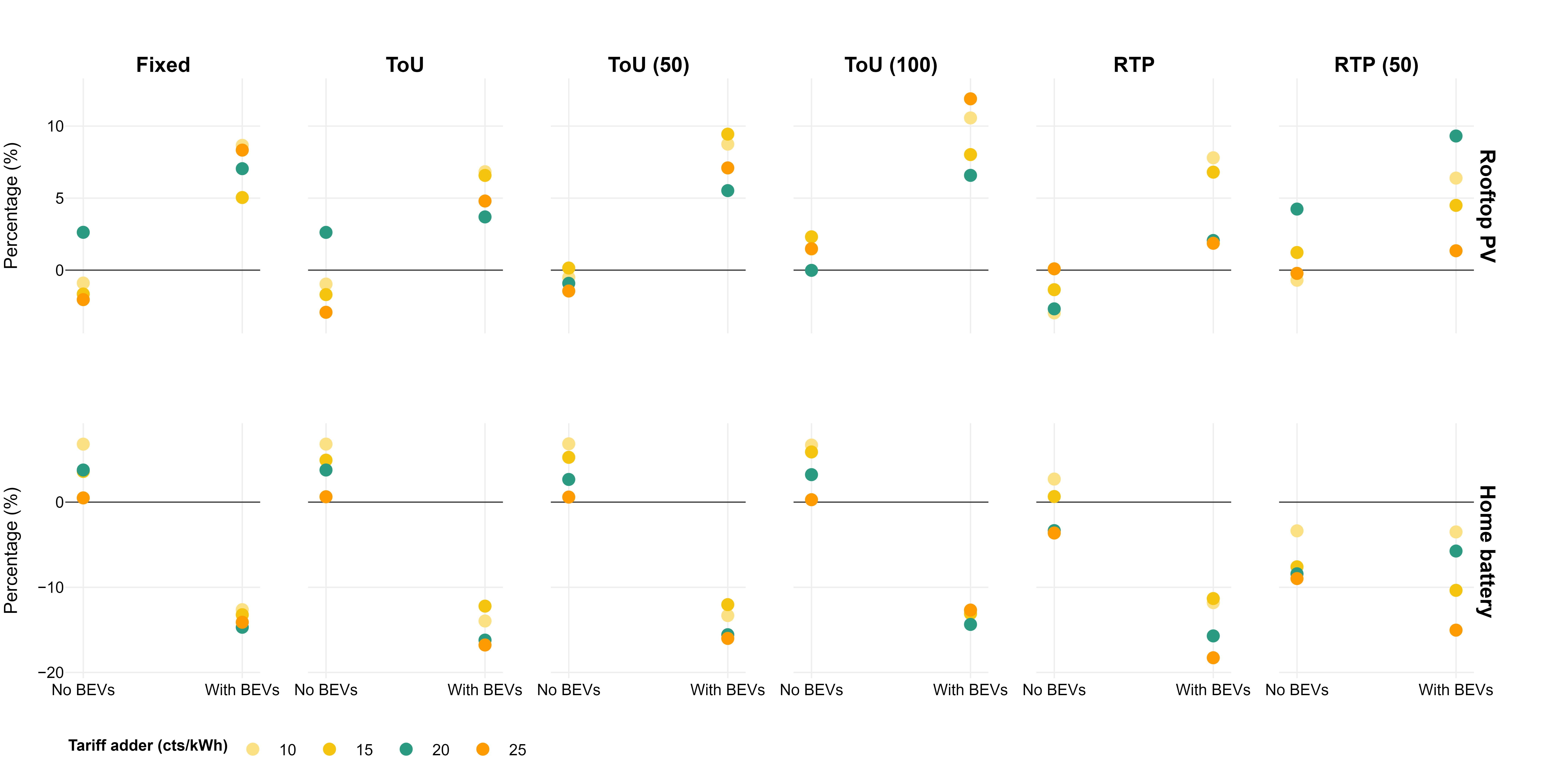}
    \caption{Deviation of the rooftop PV capacity (upper panel) and the home battery capacity (lower panel) between the central planner optimisation problem and the prosumer optimisation problem [10 million prosumers]. Note: a positive value means that the outcome in the central planner problem is larger than the outcome in the prosumer problem. Deviations are expressed relative to the optimal outcome in the prosumer bill minimisation problem.}
    \label{fig:diff_cap_10mio_relative}
\end{figure}

Deviations in electricity bills are partly explained by different rooftop PV and home battery storage capacities. In the setup \textsc{With BEVs}, the central planner problem leads to higher PV capacities in all pricing schemes (see Figure~\ref{fig:diff_cap_10mio_relative}, upper panel). By and large, the difference remains below~10\% of the optimal capacity determined in the prosumer bill minimisation. Corresponding deviations expressed in absolute capacity units are provided in the Supplemental Information (see Figure~\ref{fig:diff_cap_wortp100p_10mio_absolute}). In contrast, the central planner optimal solution systematically leads to lower investments in home battery storage energy capacity in the \textsc{With BEVs} setup (see Figure~\ref{fig:diff_cap_10mio_relative}, lower panel). This difference is clear across all pricing schemes, being at least about~10\% and reaches up to~18\%.

In the \textsc{No BEVs} setup, deviations in installed capacities are qualitatively different as they do not show a systematic overinvestment in rooftop PV and underinvestment in home battery capacity from the central planner. While there is no clear pattern, the deviations are much smaller for both technologies than in the \textsc{With BEVs} setup. For rooftop PV, deviations remain below~4\%. Deviations for home battery storage never exceed 9\%. 

The systematic overinvestment in rooftop PV and underinvestment in home battery storage in the central planner problem is driven by BEV charging episodes. The central planner takes into account the overall system costs, and not just prosumer costs. In hours when the residual load is negative and prices are accordingly low, it is optimal for the central planner to charge BEVs, including when they are not parked at home. In contrast, the different retail pricing schemes assumed here do not perfectly (if at all) inform prosumers about when the electricity for charging is the cheapest. In particular, depending on the retail pricing scheme, charging away is not necessarily cheaper than charging at home for prosumers. Hence, prosumers might charge away only when it is required to fulfil their mobility needs. In turn, charging BEVs in hours when the electricity is cheapest in the central planner problem makes them less available to consume self-generated electricity (either directly or stored). In order to still fulfil the self-generation constraint, the central planner invests in bigger rooftop PV capacity and cycles the home battery storage more to supply the prosumer domestic load.  

For all previously discussed pricing schemes, deviations of the self-generation rate and electricity bill remain below a 10\% threshold. A noticeable exception is the fully time-varying real-time pricing scheme (RTP 100), where deviations prove to be much higher, in particular for the self-generation rate where they amount to up to 20 percentage points (see Figure~\ref{fig:diff_rtp100p_10mio}). The main reason is that RTP~100 leads to price spreads which are much larger than those in other retail pricing schemes and also in wholesale prices. Very high price periods create an incentive for prosumers to build rooftop PV and home storage capacities, which help to minimise the use of grid electricity in such hours. However, in sunny hours, very low retail prices in combination with the assumed feed-in tariffs make it cheaper for prosumers to simultaneously buy electricity from the grid and feed back self-generated electricity to the grid than to consume self-generated electricity directly or store it. Such consumption patterns do not arise in the central planner problem since it is agnostic to these price frictions, leading to higher self-generation rates. This phenomenon is amplified when prosumers own a BEV, which they charge with grid electricity during hours with very low prices. Likewise, during these periods, part of the self-generated electricity that is used to charge the BEV in the central planner problem is fed to the grid in the prosumer problem. This leads to even larger deviations in the self-generation rate. Note that these deviations depend on the assumption that the prosumer battery cannot be charged from the grid, and also on the parametrisation and design of the feed-in tariff. Such deviations are thus likely to decrease with lower or dynamic feed-in tariffs.

\begin{figure}[!ht]
    \centering
    \includegraphics[width=0.47\linewidth]{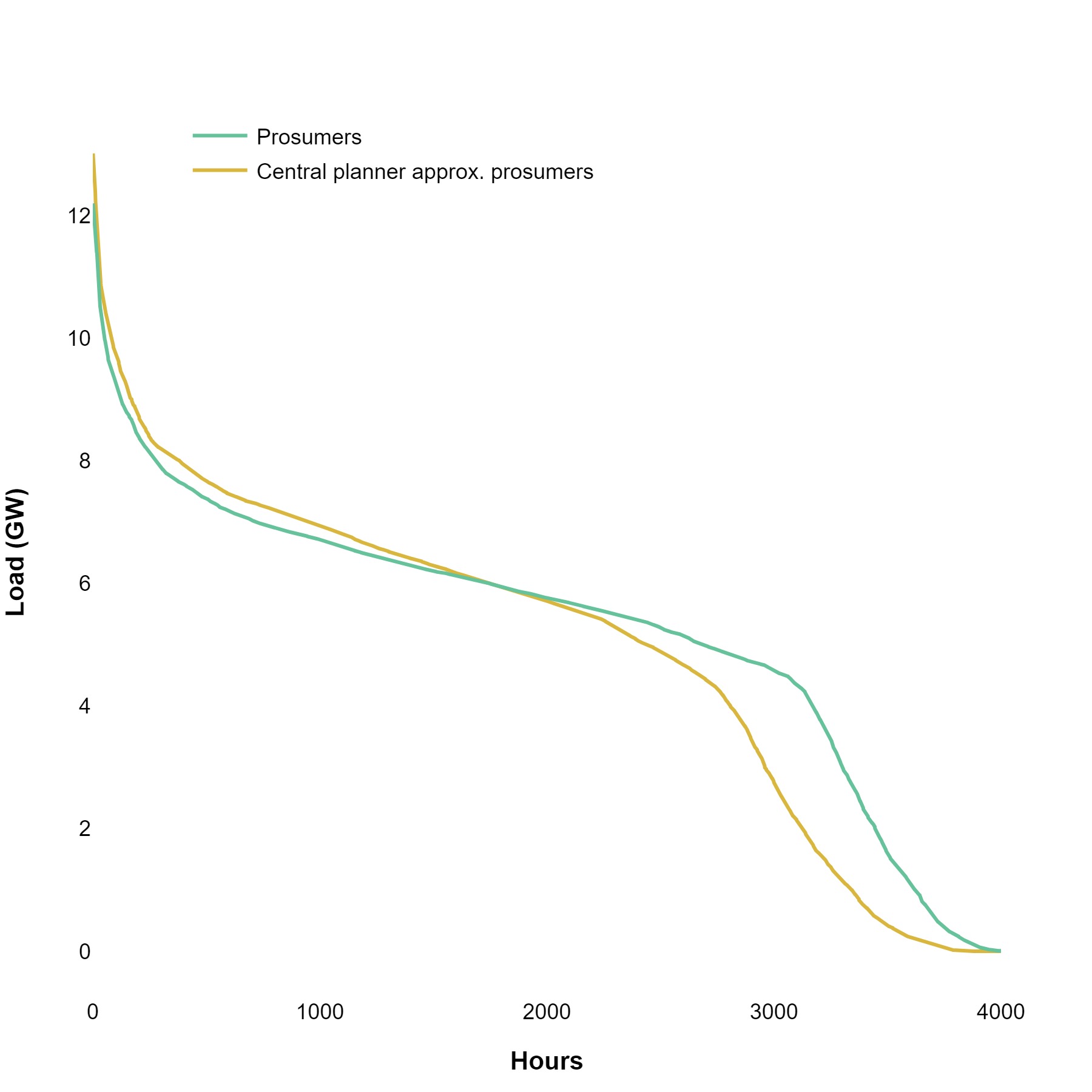}
    \includegraphics[width=0.47\linewidth]{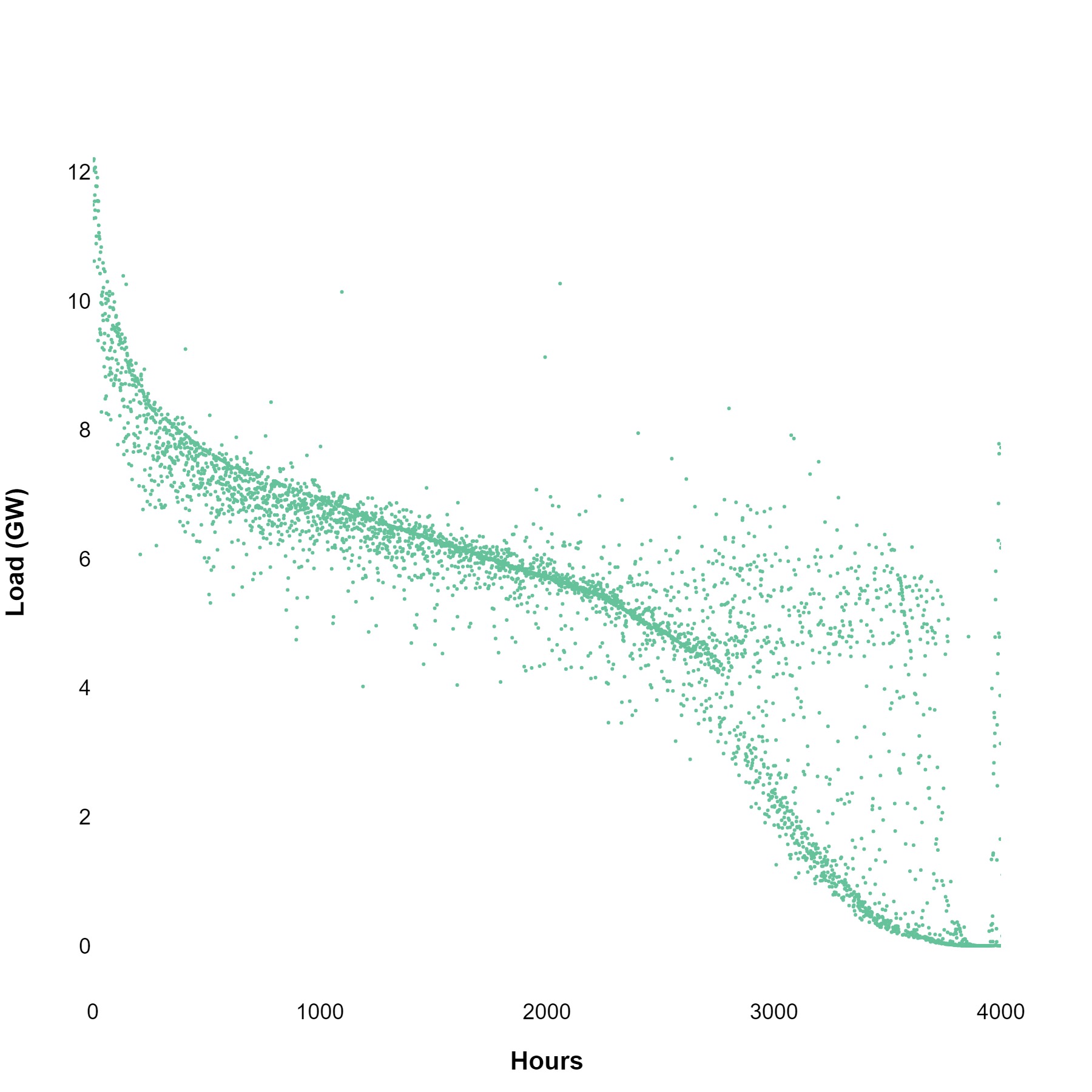}
    \caption{Rooftop PV electricity supplying prosumer load [\textsc{With BEVs}, time-invariant pricing with a tariff adder of 20 \euro cts/kWh, 10 million prosumers, first 4,000 hours]. Note: on the left panel, time series are independently sorted in decreasing order, meaning that the x-axis values do not necessarily correspond to the same hour of the year in both cases; on the right panel, values are sorted according to the central planner duration curve from the left panel.}
    \label{fig:dispatch_pv_to_p}
\end{figure}

Beyond aggregate values, comparing how technologies are temporally operated in the central planner problem with prosumer constraints and in the prosumer bill minimisation problem sheds light on both similarities and deviations. When sorted in decreasing order, duration curves of self-generated rooftop PV electricity supplying the prosumer load follow a common pattern (see left panel in Figure~\ref{fig:dispatch_pv_to_p}). Note that both curves are sorted separately in decreasing order. From hour zero until the two lines cross, directly consumed rooftop PV generation in the central planner case lies slightly above the prosumer line, which is likely driven by the larger rooftop PV capacity in the central planner case. After the two lines cross, rooftop PV supplies a smaller part of the domestic prosumer load. This can be explained by the fact that the central planner shifts more self-generated rooftop PV electricity to the home battery storage and feeds back more into the grid than the prosumer does. This is driven by the prosumer not (fully) facing wholesale prices. When sorted according to the duration curve of the central planner (see right panel in Figure~\ref{fig:dispatch_pv_to_p}), the rooftop PV supply to the load in the prosumer problem exhibits a largely similar pattern. Some temporal deviations exist, but these should not be overinterpreted, as they do not quantify the extent of the temporal mismatch between both time series. Additional comparisons of temporal patterns are provided in the Supplemental Information (see Figure~\ref{fig:dispatch_pv_to_ev} to Figure~\ref{fig:dispatch_net_imports}).  

Overall, the self-generation rate, the prosumer electricity bill and capacity investments in rooftop PV and home batteries derived in the central planner problem approximating prosumers only deviate to a limited extent from the values of the respective variables in the optimal solution of the prosumer bill minimisation. By and large, deviations for the self-generation rate and for the prosumer electricity bill remain below a 10\% threshold for all pricing schemes, except for a fully time-varying real-time pricing scheme (RTP 100). For this reason, we exclude this last pricing scheme from our subsequent analyses. 

\subsection{Impacts of approximating prosumers on the central planner optimal decisions} 

\begin{figure}[!ht]
    \centering
    \includegraphics[width=0.95\linewidth]{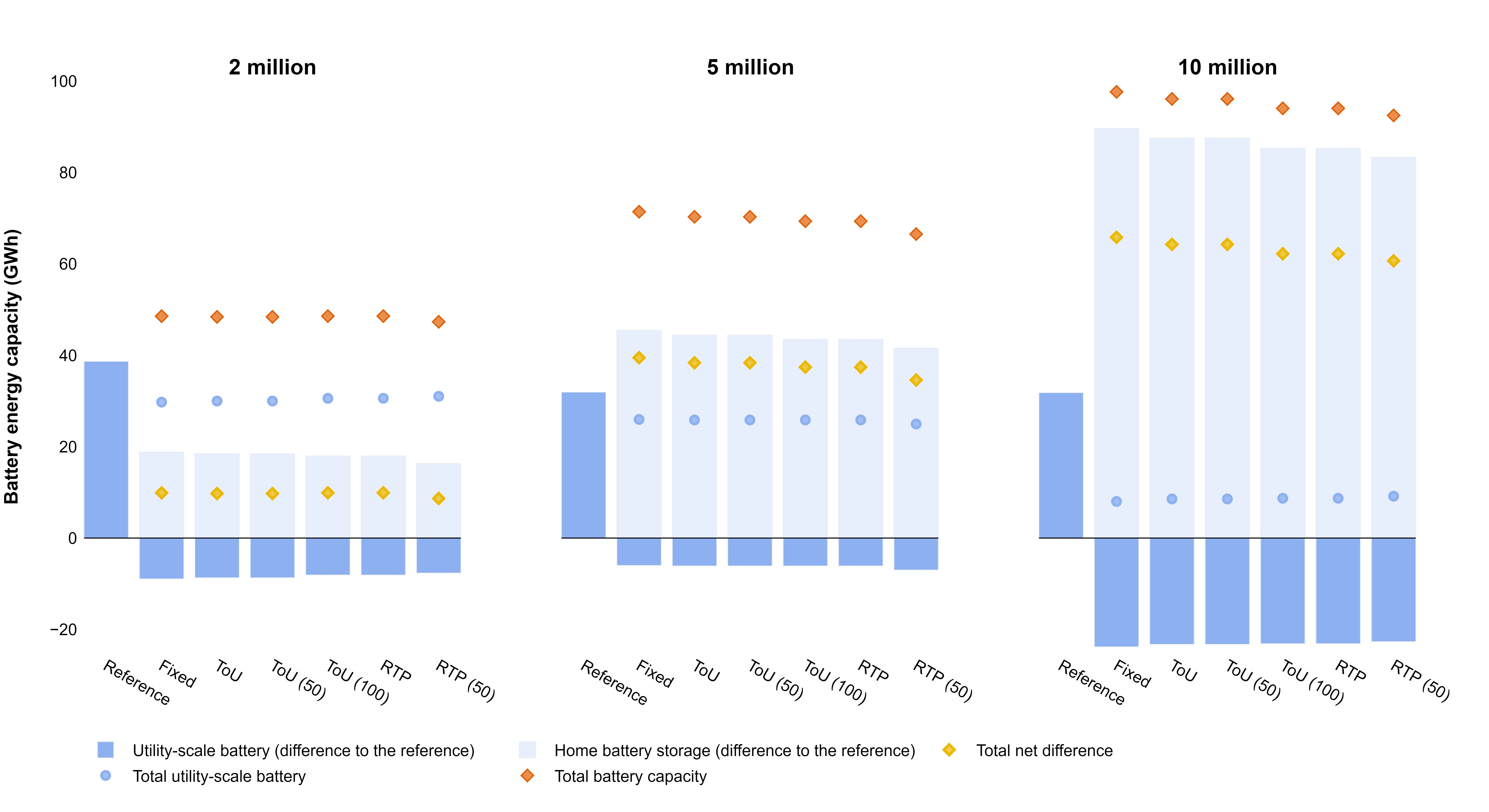}
    \caption{Battery energy capacity (in GWh) for different numbers of prosumers in the reference and scenarios [\textsc{With BEVs}, tariff adder of 20 euro cts/kWh]. Note: while bars for the reference shows absolute values, all other bars indicate differences to the respective reference.}
    \label{fig:battery_cap_withev}
\end{figure}

Approximating prosumer bill minimisation within a central planner problem as previously described impacts optimal generation and storage capacities and their hourly use. We find the largest differences for short-duration storage. Indeed, as shown in Figure~\ref{fig:battery_cap_withev}, home batteries are no perfect substitutes for utility-scale batteries. For instance, in the \textsc{With BEVs} setup with~5 or 10~million prosumers, the home battery energy capacity alone exceeds the optimal utility-scale battery capacity of the reference. Still, the optimal solution includes utility-scale batteries on top of home battery storage. 

The imperfect substitution between home and utility-scale batteries leads to a larger total battery energy capacity in the central planner optimal solution when approximating prosumers (contrast orange rhombuses with dark blue bars of the reference). In the \textsc{With BEVs} setup, the additionally installed energy capacity (yellow rhombuses) represents up to 200\% of the installed energy capacity in the reference. When considering \textsc{No BEVs}, the battery ``overcapacity'' is less pronounced across all pricing schemes but can amount to around the same energy capacity as installed in the reference (see Figure~\ref{fig:battery_cap_noev}). 

Except for battery storage, the overall capacity mix does not differ much between the two problems (see Figure~\ref{fig:generation_storage_mix_noev} and Figure~\ref{fig:generation_storage_mix_withev}). The overall installed PV capacity~-~encompassing both rooftop and utility-scale PV~-~increases slightly when approximating prosumers (see Figure~\ref{fig:pv_capacity_noev} and Figure~\ref{fig:pv_capacity_withev}). Part of this increase may be explained by an overestimation of the optimal rooftop PV capacity in the central planner problem, as underlined in Section~\ref{sec:checks}. Overall, there is an almost perfect substitution between the additionally installed rooftop PV capacity and the utility-scale PV capacity.

\begin{figure}[!ht]
    \centering
    \includegraphics[width=1.0\linewidth]{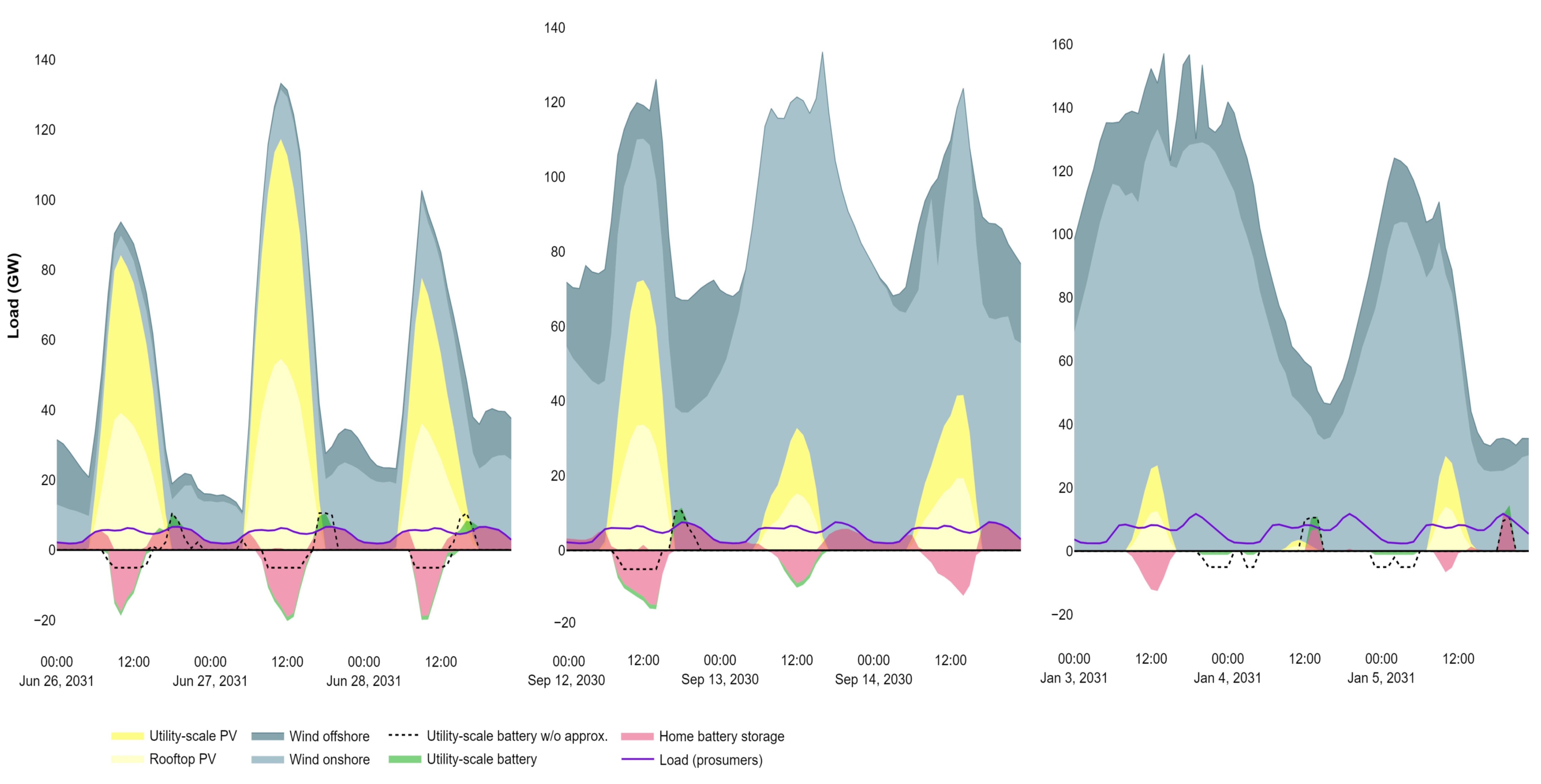}
    \caption{Power generation and battery operation for selected three-day periods [\textsc{With BEVs}, time-invariant pricing with a tariff adder of 20~\euro~cts/kWh, 10~million prosumers]. Note: operation of utility-scale battery (in green) and home battery storage (in red) are stacked, except for utility-scale battery without approximation (dotted line).}
    \label{fig:dispatch_battery_withev_june_sept_jan}
\end{figure}

Even in settings where approximating prosumers leads to overestimate installed home battery storage capacity, the magnitude of this overestimation alone cannot explain the differences observed between the optimal investments in a model with no approximation of prosumer bill minimisation compared to a model with approximation. The discrepancy in overall installed battery storage energy capacity is driven by the special characteristics of home battery storage operation. Being located behind the meter, home batteries are mainly operated to integrate self-generated electricity from rooftop PV panels, while utility-scale batteries are operated to minimise the overall dispatch costs in the system, which also includes integrating wind power. In hours when solar resources are abundant, home batteries and utility-scale batteries charge at the same time. Yet, they do not discharge in a similar fashion: home batteries discharge gradually to match the prosumer load during the evening and night when rooftop PV panels do not generate electricity, while utility-scale batteries discharge fully and rather quickly to meet the evening load peak in the overall system (Figure~\ref{fig:dispatch_battery_withev_june_sept_jan}, left panel). In hours when generation not only from solar but also from wind power is abundant, utility-scale batteries do not operate while home batteries keep charging self-generated rooftop electricity and discharging it during the night to serve the prosumer load (Figure~\ref{fig:dispatch_battery_withev_june_sept_jan}, central panel). Lastly, when solar availability is low but wind resources are abundant, utility-scale batteries charge overnight and discharge during the day. Home batteries cannot operate in such a fashion due to the absence of solar PV during these hours. Since part of utility-scale batteries are displaced by home batteries, utility-scale batteries can integrate less variable renewable electricity during winter periods (Figure~\ref{fig:dispatch_battery_withev_june_sept_jan}, right panel).

Representing prosumers in central planner optimisation problems hence leads to differences in optimal investment and dispatch of utility-scale batteries, which are related to their imperfect substitutability with home batteries. This imperfect substitutability derives from our modelling assumption - which largely reflects the current situation in Germany - that home batteries do not directly interact with the grid. They can substitute utility-scale battery storage only to some degree but cannot balance supply and demand in times of low solar and high wind availability. Such operations correspond to a large extent to how PV-battery systems are nowadays used, in the absence of financial incentives and technological means for households to operate their batteries in a system-oriented way. 

Our results hence suggest that the overall need for utility-scale battery storage in a system with prosumers is lower than what standard energy system models would find optimal. However, one cannot consider home battery systems as complete substitutes for utility-scale batteries. This imperfect substitutability leads to higher overall (utility-scale and rooftop) battery storage capacity investments than suggested by energy system models which do not approximate prosumer bill minimisation. 

\section{Discussion and conclusion}
\label{sec:discussion}

This work shows that approximating prosumers that are not - or only imperfectly - exposed to system prices in energy models leads to different results than energy models that neglect them. We find deviating solar PV and battery storage capacities and different charging patterns of stationary and vehicle batteries. In particular, we illustrate that PV-batteries are imperfect substitutes for utility-scale batteries, which corroborates and extends previous findings\autocite{say2020degrees, schick2020role, sarfarazi2023improving}. More precisely, we find that optimal battery capacity needs might be up to 200\% higher than what central planner models without prosumer constraints find.

As with any model analysis, the results are valid in the context of the particular modelling and input data choices that we made. In particular, we model prosumers in a simplified way since they are all assumed to have the same electricity demand and face the same weather conditions. In reality, spatial differences in local solar irradiation might induce differentiated behaviours and capacity investments that are not modelled in this work. A valuable extension of our work would be to test how results change when considering differentiated load and solar PV availability profiles for prosumers in a model setting with a higher geographical resolution. A potential next step would be to explicitly represent distribution and transmission grids in our model setup and investigate their interactions with prosumers. Such work could complement other analyses of distributed PV and storage in distribution\autocite{rahdan2024distributed} or transmission\autocite{kirchem2025solar} grids.

Next, our results depend on the assumed time series of BEV driving and charging availability. Prosumers likely constitute a specific group of agents, typically owning a house and able and willing to install rooftop PV panels, home battery storage and to buy electric cars. Hence, it is also likely that they live in specific geographic areas, such as suburban and rural areas, which is a well-known determinant of car driving patterns\autocite{berrill2024comparing}. The effects of alternative BEV-related assumptions could be explored in future research. Similarly, it would be valuable to include heat pumps and associated heat storage in the prosumer technology portfolio as they also represent a potentially large as well as flexible load, even though the heat demand seasonal pattern typically does not match rooftop PV availability very well.   

For simplicity, we abstract from modelling a larger penetration of non-prosumer electric cars and from Germany's interconnection with neighbouring countries. Electric cars could provide additional short-term flexibility\autocite{gueret2024impacts}, while the European interconnections could provide longer-duration flexibility\autocite{roth2023geographical}. This could reduce the need for utility-scale battery storage and dampen the ``overcapacity'' effect that we observe in our analysis. 

Despite some limitations, we believe our results provide useful insights into the pitfalls that come along not representing prosumers in energy system models. In particular, home battery storage systems cannot perfectly substitute utility-scale batteries, unless financial incentives and technological means are implemented for households to align their storage operation with the system's needs. This would require a quick and efficient rollout of smart meters as well as access to dynamic electricity retail tariffs. As long as electricity retail prices do not reflect wholesale price signals, it seems necessary to take prosumer decisions into account in energy systems models, as prosumers importance will keep increasing role in future energy systems. The method we present in this work allows to do so. Sanity checks confirm that this approach leads to solutions that closely approximate prosumer decisions for most of the considered pricing schemes. As this method is generally straightforward, it is well-suited to be implemented in other energy models, even large ones.

\singlespacing
\section*{Author contributions}

Conceptualisation: A.G., F.S., W.S.;
Methodology: A.G., F.S., W.S.;
Software: A.G., F.S.;
Formal analysis: A.G.;
Investigation: A.G.;
Data Curation: A.G., F.S.;
Writing~-~Original Draft: A.G.;
Writing~-~Review \& Editing: A.G., F.S., W.S.;
Visualisation: A.G.;
Funding acquisition: W.S. 

\section*{Data and code availability}

All data and codes are available under permissive licenses in a dedicated GitLab repository public repository at \href{https://gitlab.com/diw-evu/projects/prosumer-sector-coupling}{https://gitlab.com/diw-evu/projects/prosumer-sector-coupling}. The general repository of the DIETERjl model can also be found on GitLab at \href{https://gitlab.com/diw-evu/dieter_public/DIETERjl}{https://gitlab.com/diw-evu/dieter\_public/DIETERjl}. A complete installation guide and documentation of \textit{emobpy} can be found at \href{https://pypi.org/project/emobpy/}{https://pypi.org/project/emobpy/}. 

\section*{Acknowledgments}
We thank the participants of a DIW Berlin Brown Bag Seminar for valuable comments on a previous draft. We acknowledge a research grant by the German Federal Ministry of Education and Research via the ``Ariadne'' projects (Fkz 03SFK5NO \& 03SFK5NO-2).

\section*{Declaration of interests}

The authors declare no competing interests.

\printbibliography

@techreport{acer2024retail,
    title       = {Energy retail - Active consumer participation is key to driving the energy transition: how can it happen?},
    institution = {ACER-CEER},
    year        = {2024},
    pages       = {68},
    url         = {https://www.acer.europa.eu/sites/default/files/documents/Publications/ACER-CEER_2024_MMR_Retail.pdf}
}

@article{agnew2015effect,
  title     = {Effect of residential solar and storage on centralized electricity supply systems},
  author    = {Agnew, Scott and Dargusch, Paul},
  journal   = {Nature Climate Change},
  volume    = {5},
  number    = {4},
  pages     = {315--318},
  year      = {2015},
  publisher = {Nature Publishing Group UK London},
  doi       = {10.1038/nclimate2523}
}

@article{antonini2024weather,
  title     = {Weather-and climate-driven power supply and demand time series for power and energy system analyses},
  author    = {Antonini, Enrico GA and Di Bella, Alice and Savelli, Iacopo and Drouet, Laurent and Tavoni, Massimo},
  journal   = {Scientific Data},
  volume    = {11},
  number    = {1},
  pages     = {1--10},
  year      = {2024},
  publisher = {Nature Publishing Group}, 
  doi       = {10.1038/s41597-024-04129-8}
}

@article{berrill2024comparing,
  title={Comparing urban form influences on travel distance, car ownership, and mode choice},
  author={Berrill, Peter and Nachtigall, Florian and Javaid, Aneeque and Milojevic-Dupont, Nikola and Wagner, Felix and Creutzig, Felix},
  journal={Transportation Research Part D: Transport and Environment},
  volume={128},
  pages={104087},
  year={2024},
  publisher={Elsevier},
  doi={10.1016/j.trd.2024.104087}
}

@article{breyer2018solar,
  title     = {Solar photovoltaics demand for the global energy transition in the power sector},
  author    ={Breyer, Christian and Bogdanov, Dmitrii and Aghahosseini, Arman and Gulagi, Ashish and Child, Michael and Oyewo, Ayobami Solomon and Farfan, Javier and Sadovskaia, Kristina and Vainikka, Pasi},
  journal   = {Progress in Photovoltaics: Research and Applications},
  volume    = {26},
  number    = {8},
  pages     = {505--523},
  year      = {2018},
  publisher = {Wiley Online Library}, 
  doi       = {10.1002/pip.2950}
}

@techreport{ceer2023res,
  title        = {Status Review of Renewable Support Schemes in Europe for 2020 and 2021},
  author       = {{Council of European Energy Regulators (CEER)}},
  institution  = {Council of European Energy Regulators},
  year         = {2023},
  month        = {9},
  number       = {C22-RES-80-04},
  url          = {https://www.ceer.eu/publication/status-review-of-renewable-support-schemes-in-europe-for-2020-and-2021/},
}

@techreport{danish_energy_agency_technology_2025,
  author      = {{Danish Energy Agency}},
  title       = {Technology Data for Generation of Electricity and District Heating},
  institution = {Danish Energy Agency},
  year        = {2025},
  type        = {Technical Report},
  url         = {https://ens.dk/en/analyses-and-statistics/technology-data-generation-electricity-and-district-heating},
}

@article{fares2017impacts,
  title     = {The impacts of storing solar energy in the home to reduce reliance on the utility},
  author    = {Fares, Robert L and Webber, Michael E},
  journal   = {Nature Energy},
  volume    = {2},
  number    = {2},
  pages     = {1--10},
  year      = {2017},
  publisher = {Nature Publishing Group}, 
  doi       = {10.1038/nenergy.2017.1}
}

@book{gabriel2013complementarity,
  title     = {Complementarity Modeling in Energy Markets},
  author    = {Gabriel, Steven A. and Conejo, Antonio J. and Fuller, J. David and Hobbs, Benjamin F. and Ruiz, Carlos},
  year      = {2013},
  publisher = {Springer},
  series    = {International Series in Operations Research \& Management Science},
  volume    = {127},
  doi       = {10.1007/978-1-4419-6123-5},
  isbn      = {978-1-4419-6123-5},
  url       = {https://link.springer.com/book/10.1007/978-1-4419-6123-5}
}

@article{gaete2021open,
  title     = {An open tool for creating battery-electric vehicle time series from empirical data, emobpy},
  author    = {Gaete-Morales, Carlos and Kramer, Hendrik and Schill, Wolf-Peter and Zerrahn, Alexander},
  journal   = {Scientific Data},
  volume    = {8},
  number    = {1},
  pages     = {152},
  year      = {2021},
  publisher = {Nature Publishing Group UK London}, 
  doi       = {10.1038/s41597-021-00932-9}
}

@article{gaete2021dieterpy,
	title   = {{DIETERpy}: a {Python} framework for the {Dispatch} and {Investment} {Evaluation} {Tool} with {Endogenous} {Renewables}},
	journal = {SoftwareX},
	author  = {Gaete-Morales, Carlos and Kittel, Martin and Roth, Alexander and Schill, Wolf-Peter},
	year    = {2021},
	volume  = {15},
	pages   = {100784},
	doi     = {10.1016/j.softx.2021.100784},
}

@article{gaetemorales2024power,
  title    = {Power sector effects of alternative options for de-fossilizing heavy-duty vehicles--{G}o electric, and charge smartly}, 
  author   = {Carlos Gaete-Morales and Julius Jöhrens and Florian Heining and Wolf-Peter Schill},
  volume   = {1},
  number   = {6},
  pages    = {100123},
  year     = {2024},
  journal  = {Cell Reports Sustainability},
  doi      = {10.1016/j.crsus.2024.100123}
}

@article{gueret2024impacts,
    title   = {Impacts of electric carsharing on a power sector with variable renewables},
    journal = {Cell Reports Sustainability},
    volume  = {1},
    number  = {11},
    pages   = {100241},
    year    = {2024},
    doi     = {10.1016/j.crsus.2024.100241},
    author  = {Adeline Guéret and Wolf-Peter Schill and Carlos Gaete-Morales},
}

@article{gunther2021prosumage,
  title     = {Prosumage of solar electricity: Tariff design, capacity investments, and power sector effects},
  author    = {G{\"u}nther, Claudia and Schill, Wolf-Peter and Zerrahn, Alexander},
  journal   = {Energy Policy},
  volume    = {152},
  pages     = {112168},
  year      = {2021},
  publisher = {Elsevier},
  doi       = {10.1016/j.enpol.2021.112168}
}

@article{hoppmann2014economic,
  title      = {The economic viability of battery storage for residential solar photovoltaic systems--A review and a simulation model},
  author     = {Hoppmann, Joern and Volland, Jonas and Schmidt, Tobias S and Hoffmann, Volker H},
  journal    = {Renewable and Sustainable Energy Reviews},
  volume     = {39},
  pages      = {1101--1118},
  year       = {2014},
  publisher  = {Elsevier},
  doi        = {10.1016/j.rser.2014.07.068}
}

@article{kaschub2016solar,
  title     = {Solar energy storage in German households: profitability, load changes and flexibility},
  author    = {Kaschub, Thomas and Jochem, Patrick and Fichtner, Wolf},
  journal   = {Energy Policy},
  volume    = {98},
  pages     = {520--532},
  year      = {2016},
  publisher = {Elsevier},
  doi       = {10.1016/j.enpol.2016.09.017}
}

@article{keiner2019cost,
  title     = {Cost optimal self-consumption of PV prosumers with stationary batteries, heat pumps, thermal energy storage and electric vehicles across the world up to 2050},
  author    = {Keiner, Dominik and Ram, Manish and Barbosa, Larissa De Souza Noel Simas and Bogdanov, Dmitrii and Breyer, Christian},
  journal   = {Solar Energy},
  volume    = {185},
  pages     = {406--423},
  year      = {2019},
  publisher = {Elsevier},
  doi       = {10.1016/j.solener.2019.04.081}
}

@article{kirchem2023power,
  title    = {Power sector effects of green hydrogen production in Germany},
  author   = {Kirchem, Dana and Schill, Wolf-Peter},
  journal  = {Energy Policy},
  volume   = {182},
  pages    = {113738},
  year     = {2023},
  doi      = {10.1016/j.enpol.2023.113738}
}

@misc{kirchem2025solar,
      title={Solar prosumage under different pricing regimes: Interactions with the transmission grid}, 
      author={Dana Kirchem and Mario Kendziorski and Enno Wiebrow and Wolf-Peter Schill and Claudia Kemfert and Christian von Hirschhausen},
      year={2025},
      eprint={2502.21306},
      archivePrefix={arXiv},
      primaryClass={econ.GN},
      url={https://arxiv.org/abs/2502.21306}, 
}

@article{lang2016profitability,
  title     = {Profitability in absence of subsidies: A techno-economic analysis of rooftop photovoltaic self-consumption in residential and commercial buildings},
  author    = {Lang, Tillmann and Ammann, David and Girod, Bastien},
  journal   = {Renewable Energy},
  volume    = {87},
  pages     = {77--87},
  year      = {2016},
  publisher = {Elsevier},
  doi       = {10.1016/j.renene.2015.09.059}
}

@Book{mas_colell_microeconomic_1995,
  author={Mas-Colell, Andreu and Whinston, Michael D. and Green, Jerry R.},
  title={{Microeconomic Theory}},
  publisher={Oxford University Press},
  year=1995,
  month={12},
  volume={},
  number={9780195102680},
  series={OUP Catalogue},
  edition={},
  doi={},
  url={https://ideas.repec.org/b/oxp/obooks/9780195102680.html}
}

@article{m_jrc-idees-2021_2024,
	title = {{JRC}-{IDEES}-2021: the {Integrated} {Database} of the {European} {Energy} {System} – {Data} update and technical documentation},
	issn = {1831-9424},
	doi = {10.2760/614599},
	number = {KJ-NA-31-940-EN-N},
	author = {M, Rózsai and M, Jaxa-Rozen and R, Salvucci and P, Sikora and J, Tattini and F, Neuwahl},
	year = {2024}
}

@article{neumann_potential_2023,
	title = {The potential role of a hydrogen network in {Europe}},
	volume = {7},
	issn = {25424351},
	url = {https://linkinghub.elsevier.com/retrieve/pii/S2542435123002660},
	doi = {10.1016/j.joule.2023.06.016},
	language = {en},
	number = {8},
	urldate = {2025-04-08},
	journal = {Joule},
	author = {Neumann, Fabian and Zeyen, Elisabeth and Victoria, Marta and Brown, Tom},
	month = aug,
	year = {2023},
	pages = {1793--1817},
}

@article{rahdan2024distributed,
  title      = {Distributed photovoltaics provides key benefits for a highly renewable European energy system},
  author     = {Rahdan, Parisa and Zeyen, Elisabeth and Gallego-Castillo, Cristobal and Victoria, Marta},
  journal    = {Applied Energy},
  volume     = {360},
  pages      = {122721},
  year       = {2024},
  publisher  = {Elsevier},
  doi        = {10.1016/j.apenergy.2024.122721}
}

@article{renewables2024analysis,
  title={Analysis and Forecast to 2030},
  author={Renewables, IEA},
  journal={International Energy Agency: Paris, France},
  year={2024}
}

@article{roth2023geographical,
   title   = {Geographical balancing of wind power decreases storage needs in a 100\% renewable {European} power sector},
   journal = {iScience},
   volume  = {26},
   number  = {7},
   pages   = {107074},
   year    = {2023},
   doi     = {10.1016/j.isci.2023.107074},
   author  = {Alexander Roth and Wolf-Peter Schill}
}

@article{roth2024power,
  author   = {Roth, Alexander and Gaete-Morales, Carlos and Kirchem, Dana and Schill, Wolf-Peter},
  journal  = {Communications Earth \& Environment},
  title    = {Power sector benefits of flexible heat pumps in 2030 scenarios},
  year     = {2024},
  number   = {1},
  pages    = {718},
  volume   = {5},
  doi      = {10.1038/s43247-024-01861-2},
}

@article{ruhnau_update_2022,
	title = {Update and extension of the {When2Heat} dataset},
	language = {en},
	journal = {ZBW – Leibniz Information Centre for Economics, Kiel, Hamburg},
	author = {Ruhnau, Oliver and Muessel, Jarusch},
	year = {2022},
}

@article{sarfarazi2023improving,
  title     = {Improving energy system design with optimization models by quantifying the economic granularity gap: The case of prosumer self-consumption in Germany},
  author    = {Sarfarazi, Seyedfarzad and Sasanpour, Shima and Cao, Karl-Ki{\^e}n},
  journal   = {Energy Reports},
  volume    = {9},
  pages     = {1859--1874},
  year      = {2023},
  publisher = {Elsevier},
  doi       = {10.1016/j.egyr.2022.12.145}
}

@article{say2018coming,
  title     = {The coming disruption: The movement towards the customer renewable energy transition},
  author    = {Say, Kelvin and John, Michele and Dargaville, Roger and Wills, Raymond T},
  journal   = {Energy Policy},
  volume    = {123},
  pages     = {737--748},
  year      = {2018},
  publisher = {Elsevier}, 
  doi       = {10.1016/j.enpol.2018.09.026}
}

@article{say2020degrees,
  title     = {Degrees of displacement: The impact of household PV battery prosumage on utility generation and storage},
  author    = {Say, Kelvin and Schill, Wolf-Peter and John, Michele},
  journal   = {Applied Energy},
  volume    = {276},
  pages     = {115466},
  year      = {2020},
  publisher = {Elsevier}, 
  doi       = {10.1016/j.apenergy.2020.115466}
}

@article{schick2020role,
  title      = {Role and impact of prosumers in a sector-integrated energy system with high renewable shares},
  author     = {Schick, Christoph and Klempp, Nikolai and Hufendiek, Kai},
  journal    = {IEEE Transactions on Power Systems},
  volume     = {37},
  number     = {4},
  pages      = {3286--3298},
  year       = {2020},
  publisher  = {IEEE},
  doi        = {10.1109/TPWRS.2020.3040654}
}

@article{schill2017prosumage,
  title      = {Prosumage of solar electricity: pros, cons, and the system perspective},
  author     = {Schill, Wolf-Peter and Zerrahn, Alexander and Kunz, Friedrich},
  journal    = {Economics of Energy \& Environmental Policy},
  volume     = {6},
  number     = {1},
  pages      = {7--32},
  year       = {2017},
  publisher  = {JSTOR},
  doi        = {10.5547/2160-5890.6.1.wsch}
}

@article{schill2017decentralized,
  title     = {Decentralized solar prosumage with battery storage: System orientation required},
  author    = {Schill, Wolf-Peter and Zerrahn, Alexander and Kunz, Friedrich and Kemfert, Claudia},
  journal   = {DIW Economic Bulletin},
  volume    = {7},
  number    = {12/13},
  pages     = {141--151},
  year      = {2017},
  publisher = {Berlin: Deutsches Institut f{\"u}r Wirtschaftsforschung (DIW)}
}

@article{schill2018long-run,
  title    = {Long-run power storage requirements for high shares of renewables: Results and sensitivities},
  author   = {Schill, Wolf-Peter and Zerrahn, Alexander},
  journal  = {Renewable and Sustainable Energy Reviews},
  volume   = {83},
  pages    = {156--171},
  year     = {2018},
  doi      = {doi.org/10.1016/j.rser.2017.05.205}
}

@article{schoniger2022comes,
  title     = {What comes down must go up: Why fluctuating renewable energy does not necessarily increase electricity spot price variance in Europe},
  author    = {Sch{\"o}niger, Franziska and Morawetz, Ulrich B},
  journal   = {Energy Economics},
  volume    = {111},
  pages     = {106069},
  year      = {2022},
  publisher = {Elsevier},
  doi       = {10.1016/j.eneco.2022.106069}
}

@article{schwarz2018self,
  title     = {Self-consumption through power-to-heat and storage for enhanced PV integration in decentralised energy systems},
  author    = {Schwarz, Hannes and Schermeyer, Hans and Bertsch, Valentin and Fichtner, Wolf},
  journal   = {Solar Energy},
  volume    = {163},
  pages     = {150--161},
  year      = {2018},
  publisher = {Elsevier},
  doi       = {doi.org/10.1016/j.solener.2018.01.076}
}

@article{semmelmann2024empirical,
  title     = {Empirical field evaluation of self-consumption promoting regulation of household battery energy storage systems},
  author    = {Semmelmann, Leo and Konermann, Marie and Dietze, Daniel and Staudt, Philipp},
  journal   = {Energy Policy},
  volume    = {194},
  pages     = {114343},
  year      = {2024},
  publisher = {Elsevier}, 
  doi       = {10.1016/j.enpol.2024.114343}
}

@techreport{solar2023europe,
    title       = {EU Market Outlook for Solar Power 2023-2027},
    institution = {SolarPower Europe},
    year        = {2023}
}

@article{steinbach2024enabling,
  title     = {Enabling electric mobility: can photovoltaic and home battery systems significantly reduce grid reinforcement costs?},
  author    = {Steinbach, Sarah A and Blaschke, Maximilian J},
  journal   = {Applied Energy},
  volume    = {375},
  pages     = {124101},
  year      = {2024},
  publisher = {Elsevier},
  doi       = {10.1016/j.apenergy.2024.124101}
}

@article{stute2024assessing,
  title      = {Assessing the conditions for economic viability of dynamic electricity retail tariffs for households},
  author     = {Stute, Judith and Pelka, Sabine and K{\"u}hnbach, Matthias and Klobasa, Marian},
  journal    = {Advances in Applied Energy},
  volume     = {14},
  pages      = {100174},
  year       = {2024},
  publisher  = {Elsevier}, 
  doi        = {10.1016/j.adapen.2024.100174}
}

@misc{stromnetzberlin_netznutzer,
  author       = {{Stromnetz Berlin GmbH}},
  title        = {Netznutzer},
  year         = {2025},
  url          = {https://www.stromnetz.berlin/netz-nutzen/netznutzer},
  note         = {\href{https://www.stromnetz.berlin/netz-nutzen/netznutzer}{https://www.stromnetz.berlin/netz-nutzen/netznutzer}}
}

@article{wiese2019open,
	title = {Open {Power} {System} {Data} – {Frictionless} data for electricity system modelling},
	volume = {236},
	doi = {10.1016/j.apenergy.2018.11.097},
	journal = {Applied Energy},
	author = {Wiese, Frauke and Schlecht, Ingmar and Bunke, Wolf-Dieter and Gerbaulet, Clemens and Hirth, Lion and Jahn, Martin and Kunz, Friedrich and Lorenz, Casimir and Mühlenpfordt, Jonathan and Reimann, Juliane and Schill, Wolf-Peter},
	year = {2019},
	pages = {401--409},
}

@article{yu2018prospective,
  title     = {A prospective economic assessment of residential PV self-consumption with batteries and its systemic effects: The French case in 2030},
  author    = {Yu, Hyun Jin Julie},
  journal   = {Energy Policy},
  volume    = {113},
  pages     = {673--687},
  year      = {2018},
  publisher = {Elsevier},
  doi       = {10.1016/j.enpol.2017.11.005}
}

@article{zerrahn2017long-run,
  title  = {Long-run power storage requirements for high shares of renewables: review and a new model},
  author = {Zerrahn, Alexander and Schill, Wolf-Peter},
  journal= {Renewable and Sustainable Energy Reviews},
  year   = {2017},
  volume = {79},
  pages  = {1518--1534},
  doi    = {10.1016/j.rser.2016.11.098}
}

\appendix
\setcounter{figure}{0}
\renewcommand{\thefigure}{SI.\arabic{figure}}
\renewcommand*{\theHfigure}{\thefigure}
\setcounter{table}{0}
\renewcommand{\thetable}{SI.\arabic{table}}
\renewcommand*{\theHtable}{\thetable}
\renewcommand{\thesubsection}{SI.\arabic{subsection}}

\newpage 

\section*{Supplemental Information}

\subsection*{Supplemental Figures}

\begin{figure}[!ht]
    \centering
    \includegraphics[width=0.9\linewidth]{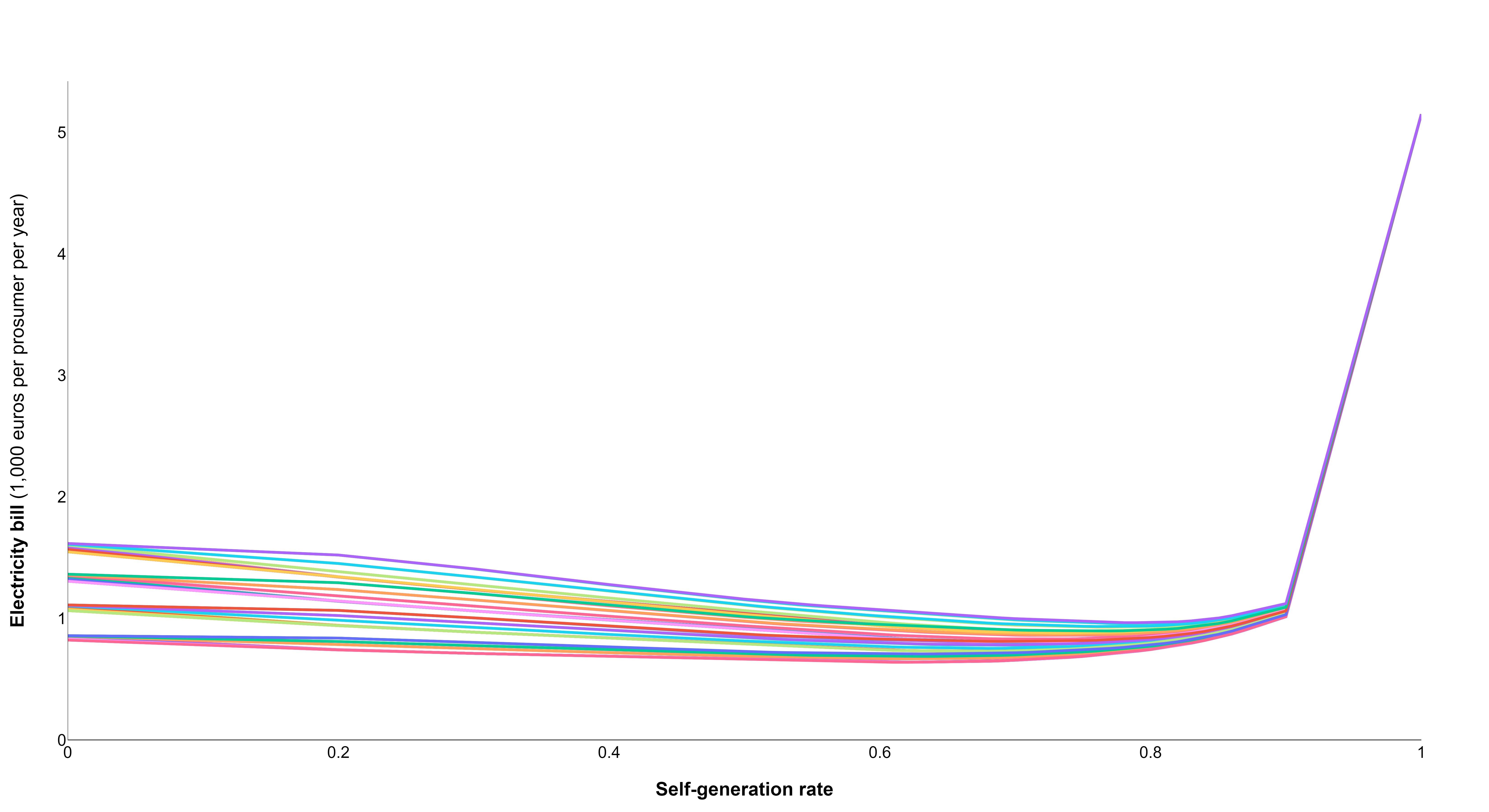}
    \caption{Prosumer total costs for all tariff schemes and number of prosumers considered}
    \label{fig:costs_total_all_noev}
\end{figure}

\begin{figure}[!ht]
    \centering
    \includegraphics[width=0.9\linewidth]{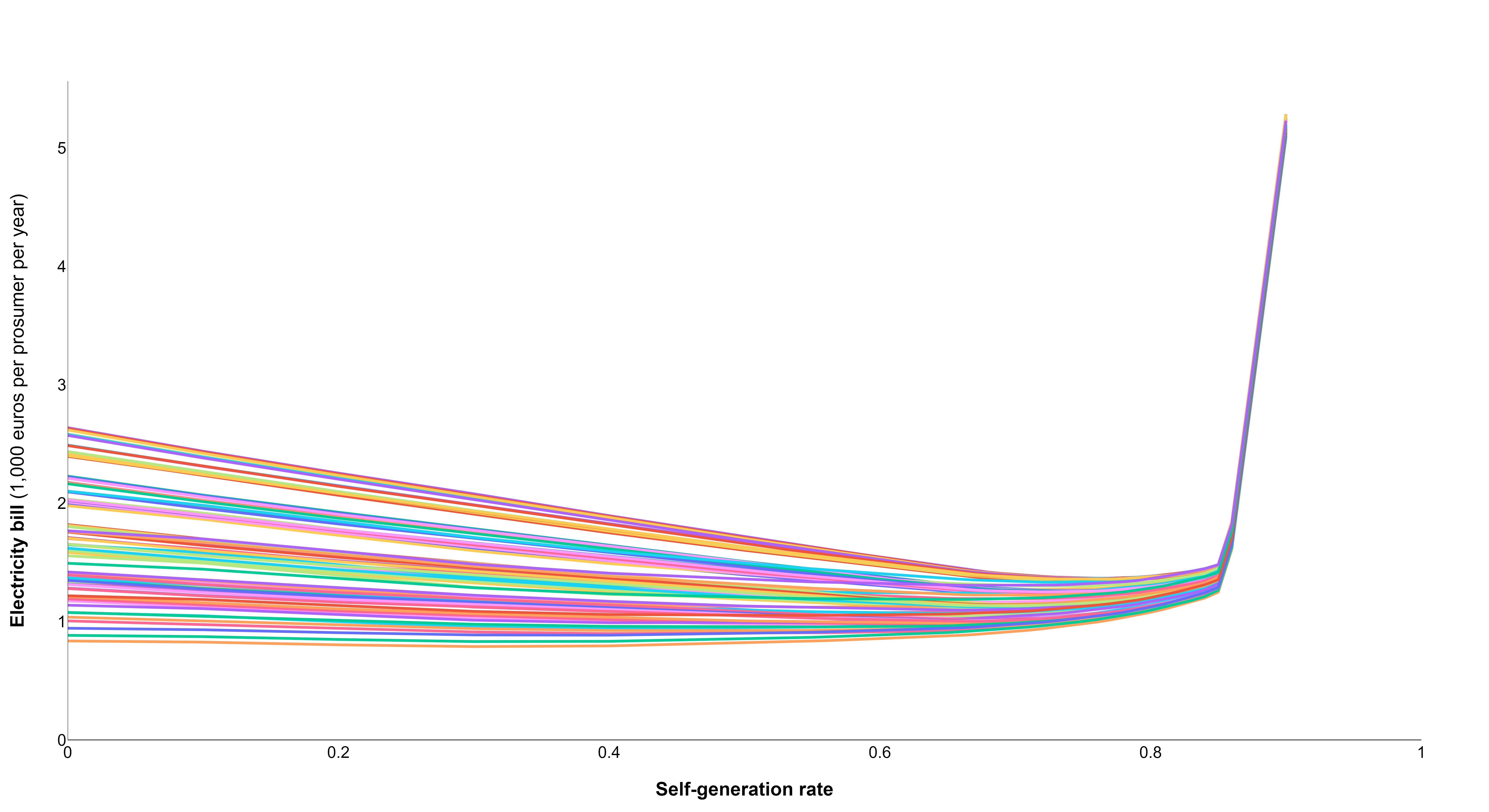}
    \caption{Prosumer total costs for all tariff schemes and number of prosumers considered}
    \label{fig:costs_total_all_withev}
\end{figure}

\begin{figure}[!ht]
    \centering
    \includegraphics[width=0.9\linewidth]{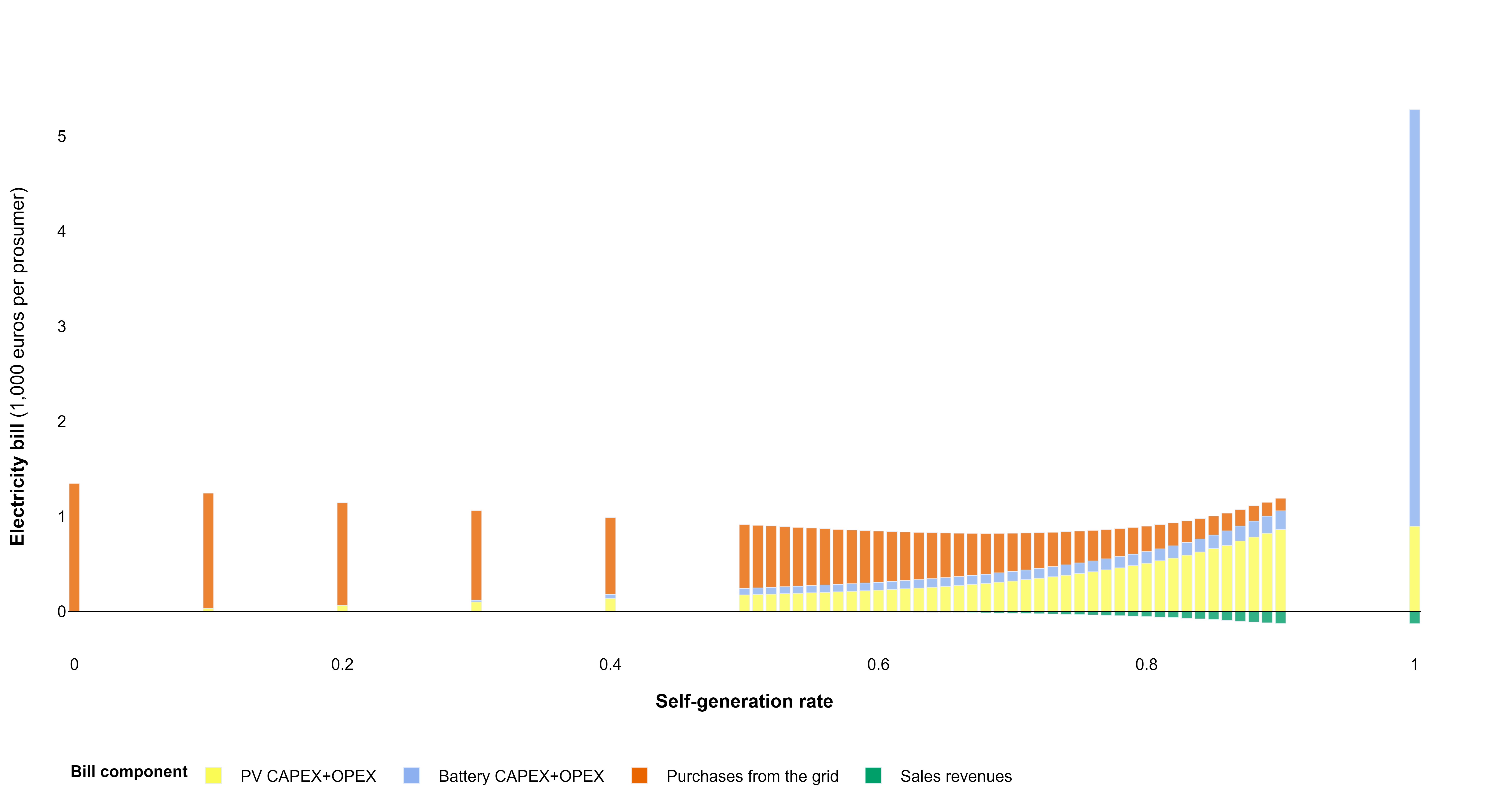}
    \caption{Prosumer bill decomposition by cost component [\textsc{No BEVs}, 10 million prosumers]}
    \label{fig:costs_decomposition_std}
\end{figure}

\clearpage

\begin{figure}[!ht]
    \centering
    \includegraphics[width=0.95\linewidth]{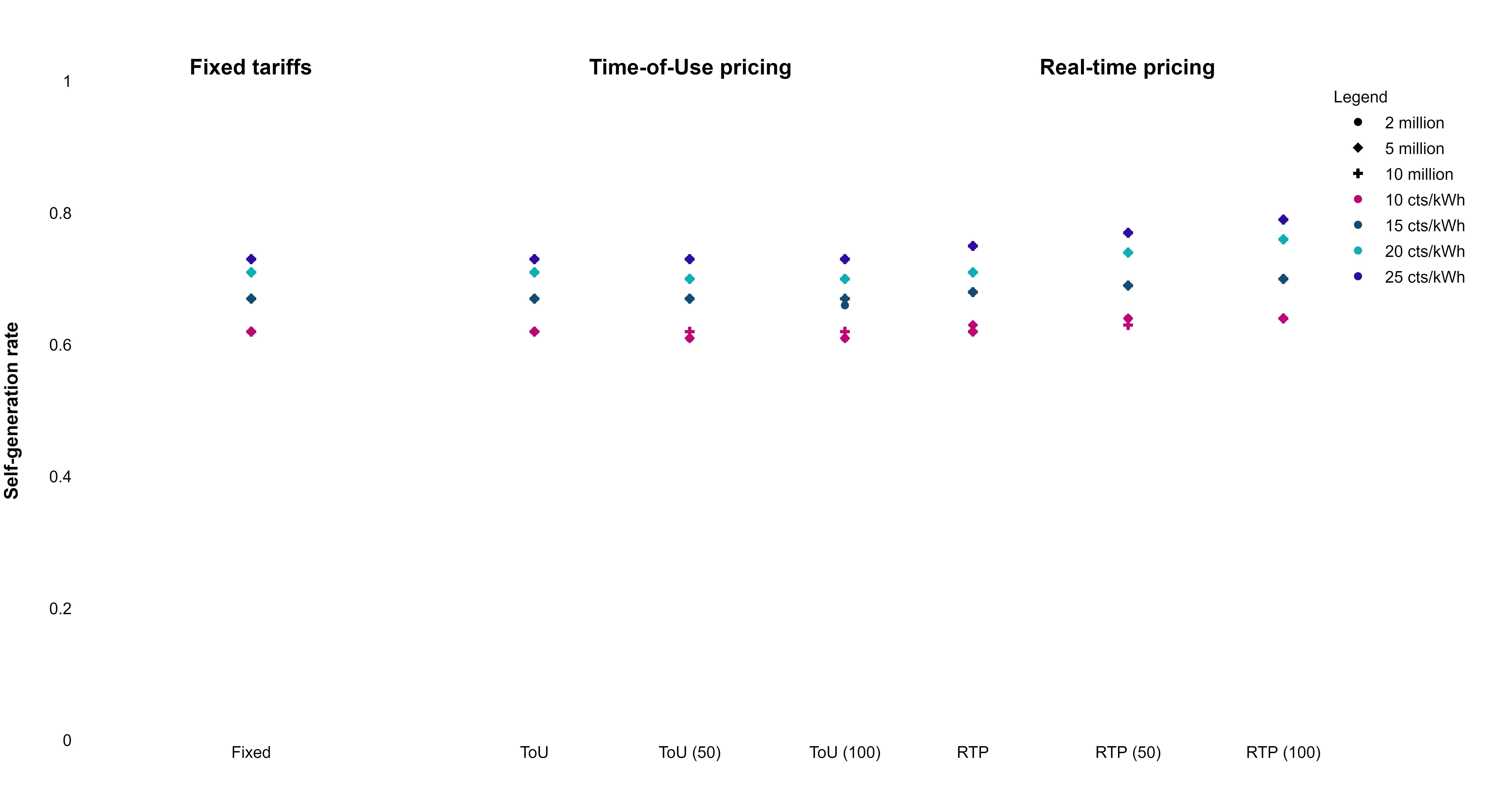}
    \caption{Bill-minimising self-generation rate in the central planner optimisation problem by pricing schemes and number of prosumers [\textsc{No BEVs}]}
    \label{fig:cost_minimising_omega_noev}
\end{figure}

\begin{figure}[!ht]
    \centering
    \includegraphics[width=0.95\linewidth]{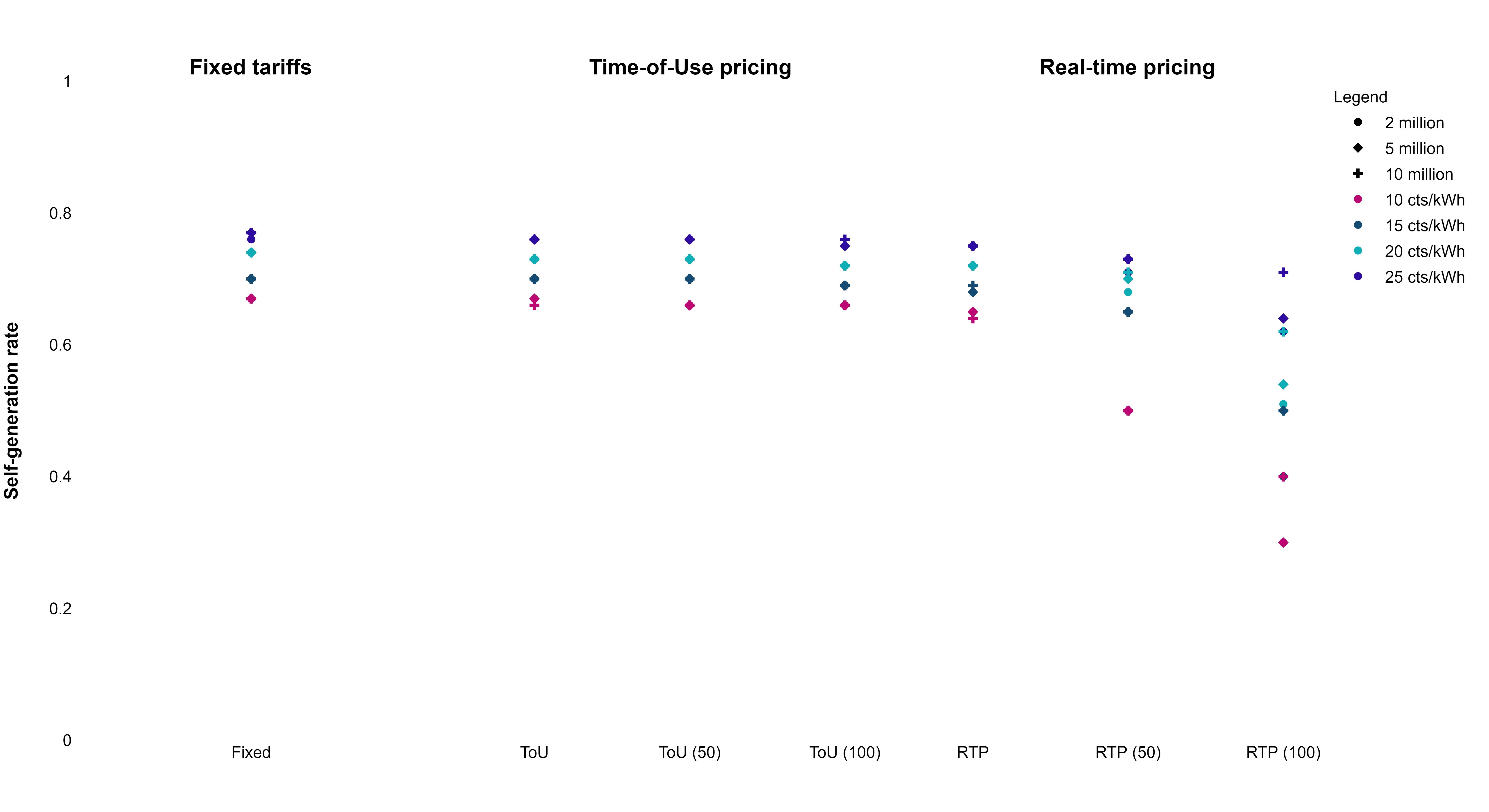}
    \caption{Bill-minimising self-generation rate in the central planner optimisation problem by pricing schemes and number of prosumers [\textsc{With BEVs}]}
    \label{fig:cost_minimising_omega_withev}
\end{figure} 

\begin{figure}[!ht]
    \centering
    \includegraphics[width=1.0\linewidth]{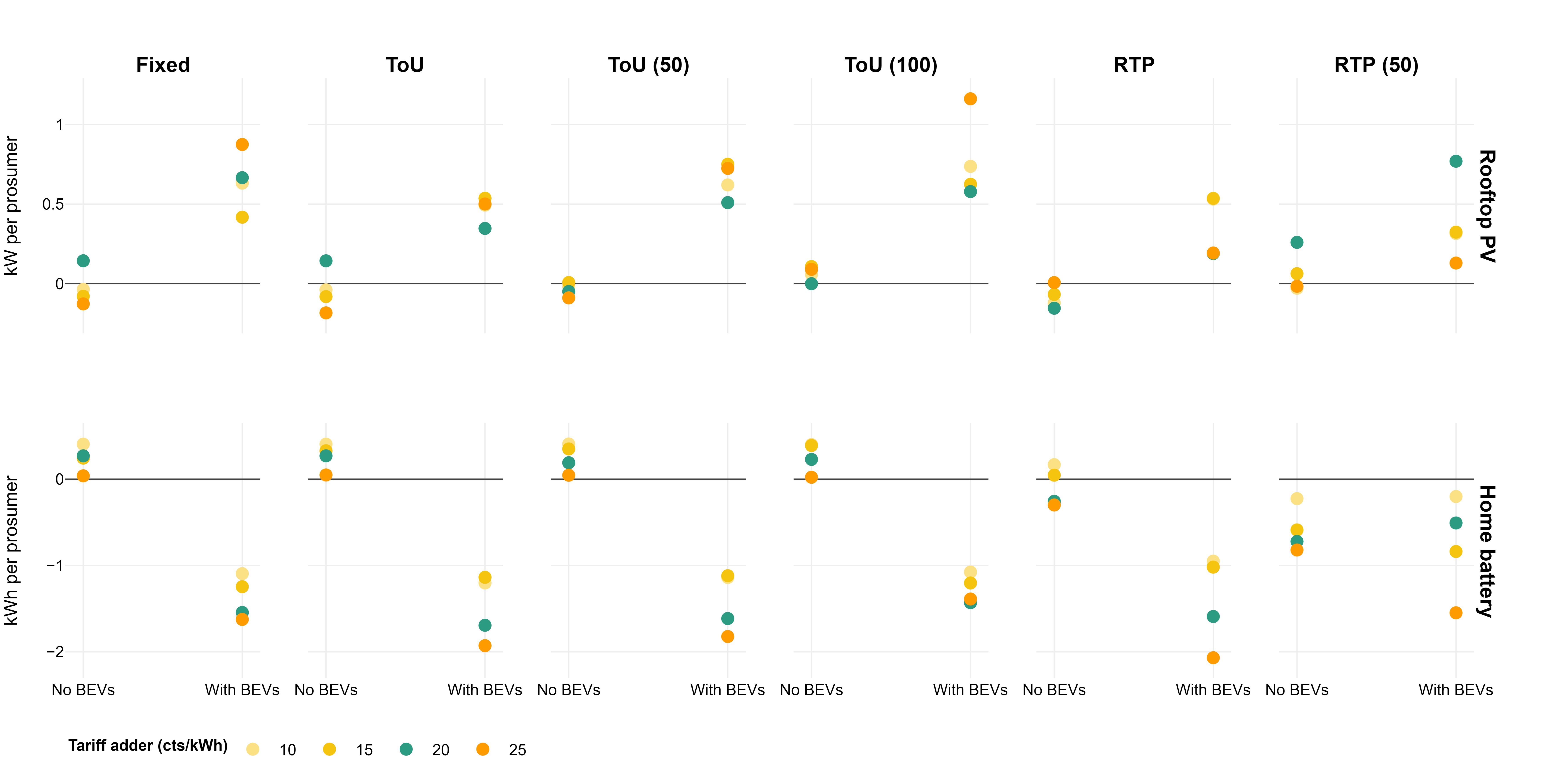}.
    \caption{Deviation of the rooftop PV capacity (upper panel) and the home battery capacity (lower panel) between the central planner optimisation problem and the prosumer optimisation problem [10 million prosumers]. Note: a positive value means that the outcome in the central planner problem is larger than the outcome in the prosumer problem. Deviations for rooftop PV are expressed in kW per prosumer and in kWh per prosumer for home batteries.}
    \label{fig:diff_cap_wortp100p_10mio_absolute}
\end{figure}

\clearpage

\begin{figure}[!ht]
    \centering
    \includegraphics[width=0.95\linewidth]{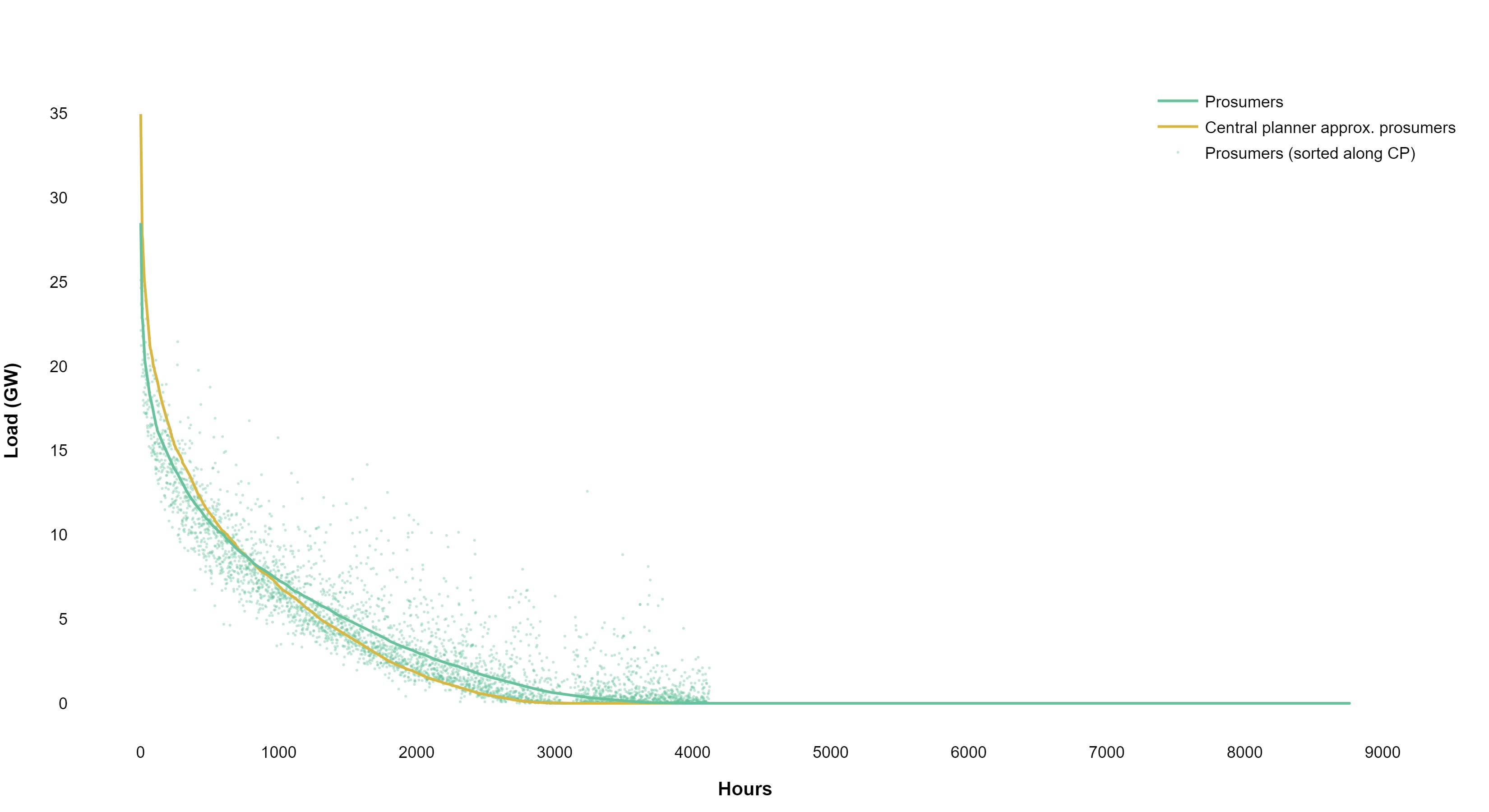}
    \caption{Self-generated electricity charging BEV at home [\textsc{With BEVs}, time-invariant pricing with a tariff adder of 20 \euro cts/kWh, 10 million prosumers]}
    \label{fig:dispatch_pv_to_ev}
\end{figure}

\begin{figure}[!ht]
    \centering
    \includegraphics[width=0.95\linewidth]{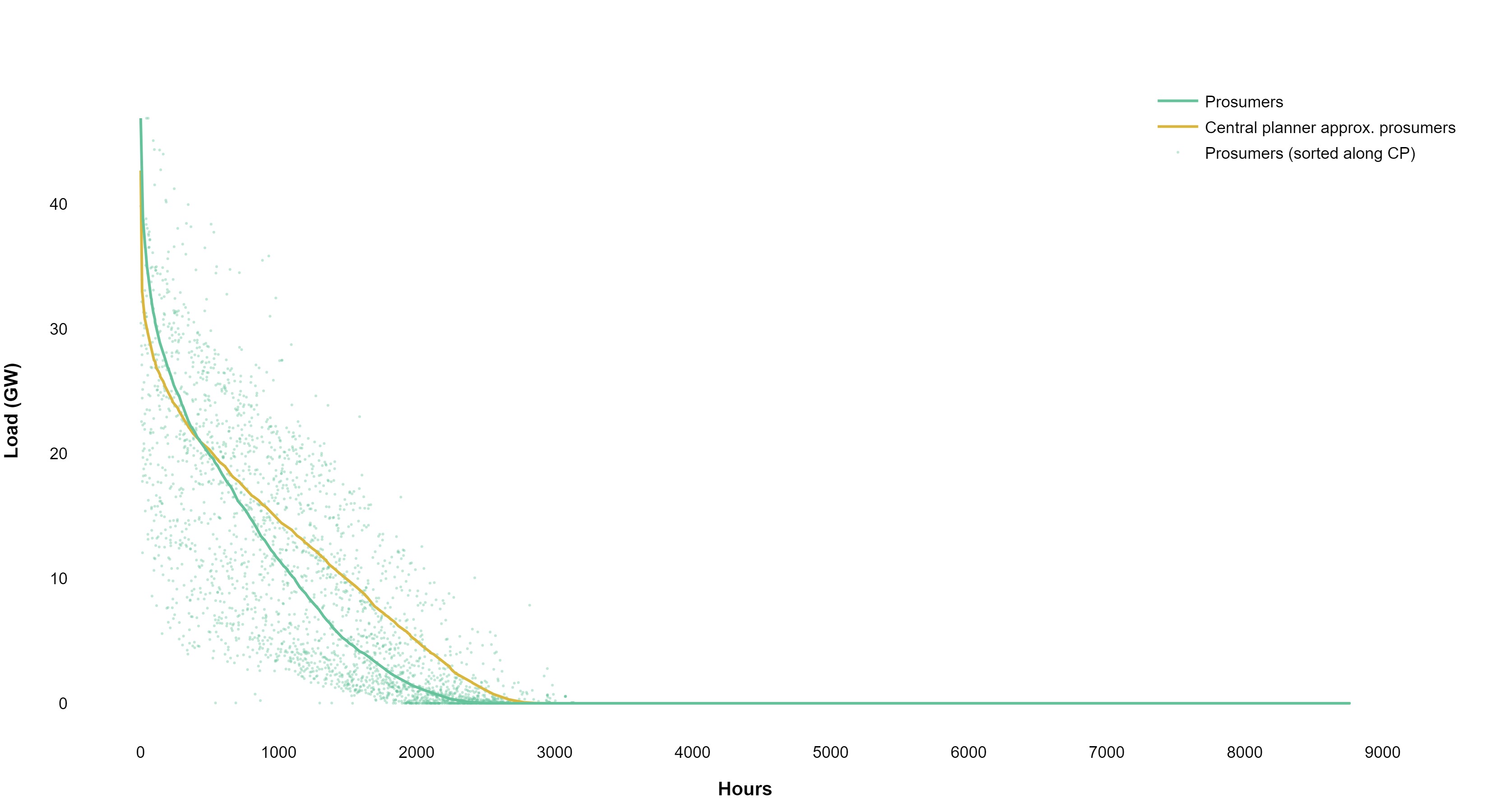}
    \caption{Self-generated electricity fed to the central electricity grid [\textsc{With BEVs}, time-invariant pricing with a tariff adder of 20 \euro cts/kWh, 10 million prosumers]}
    \label{fig:dispatch_pv_to_m}
\end{figure}

\clearpage

\begin{figure}[!ht]
    \centering
    \includegraphics[width=0.95\linewidth]{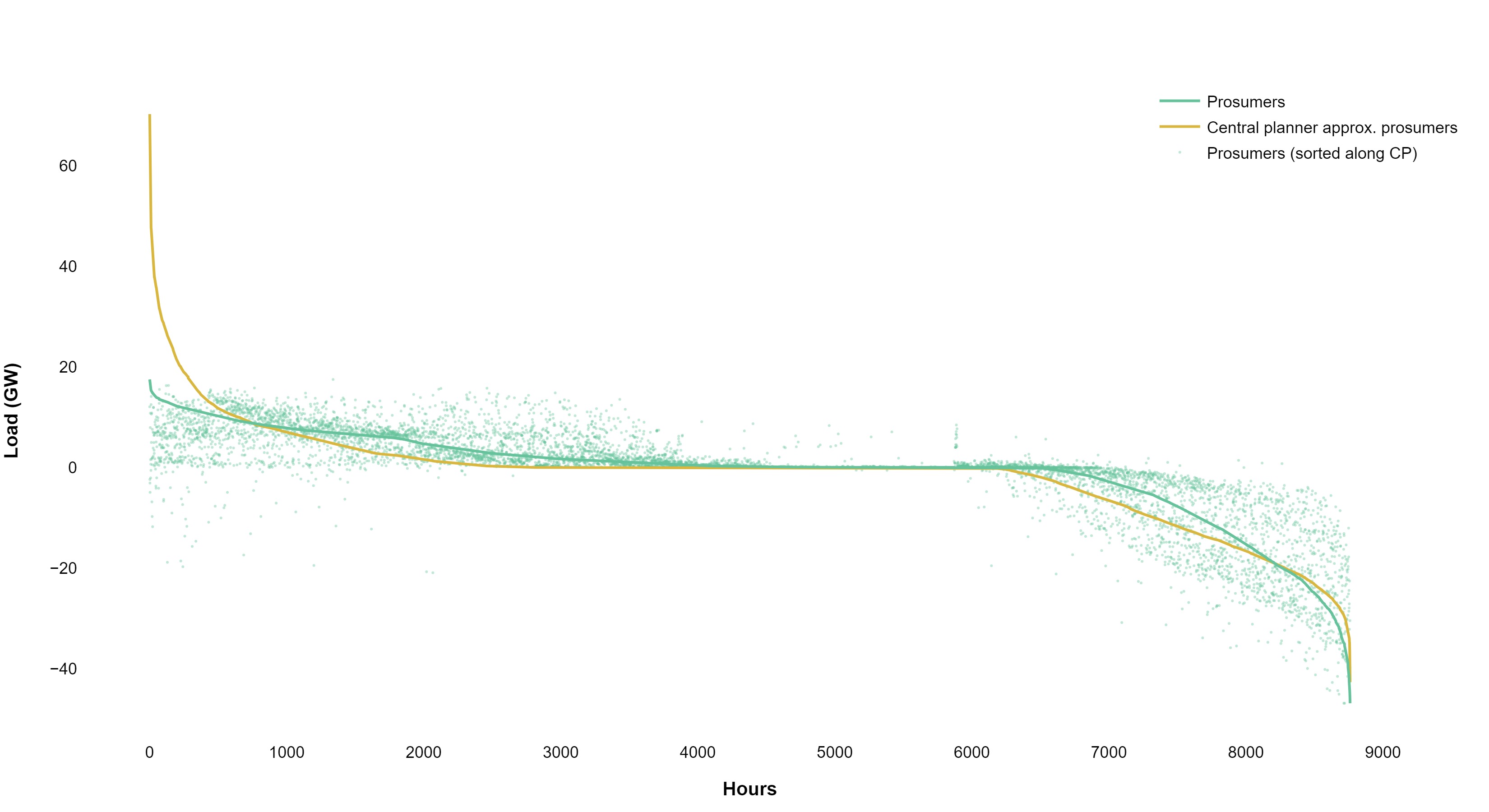}
    \caption{Prosumer net grid consumption (purchased electricity from the grid minus overall self-generated electricity fed into the grid) [\textsc{With BEVs}, time-invariant pricing with a tariff adder of 20 \euro cts/kWh, 10 million prosumers]}
    \label{fig:dispatch_net_imports}
\end{figure}

\begin{figure}[!ht]
    \centering
    \includegraphics[width=0.95\linewidth]{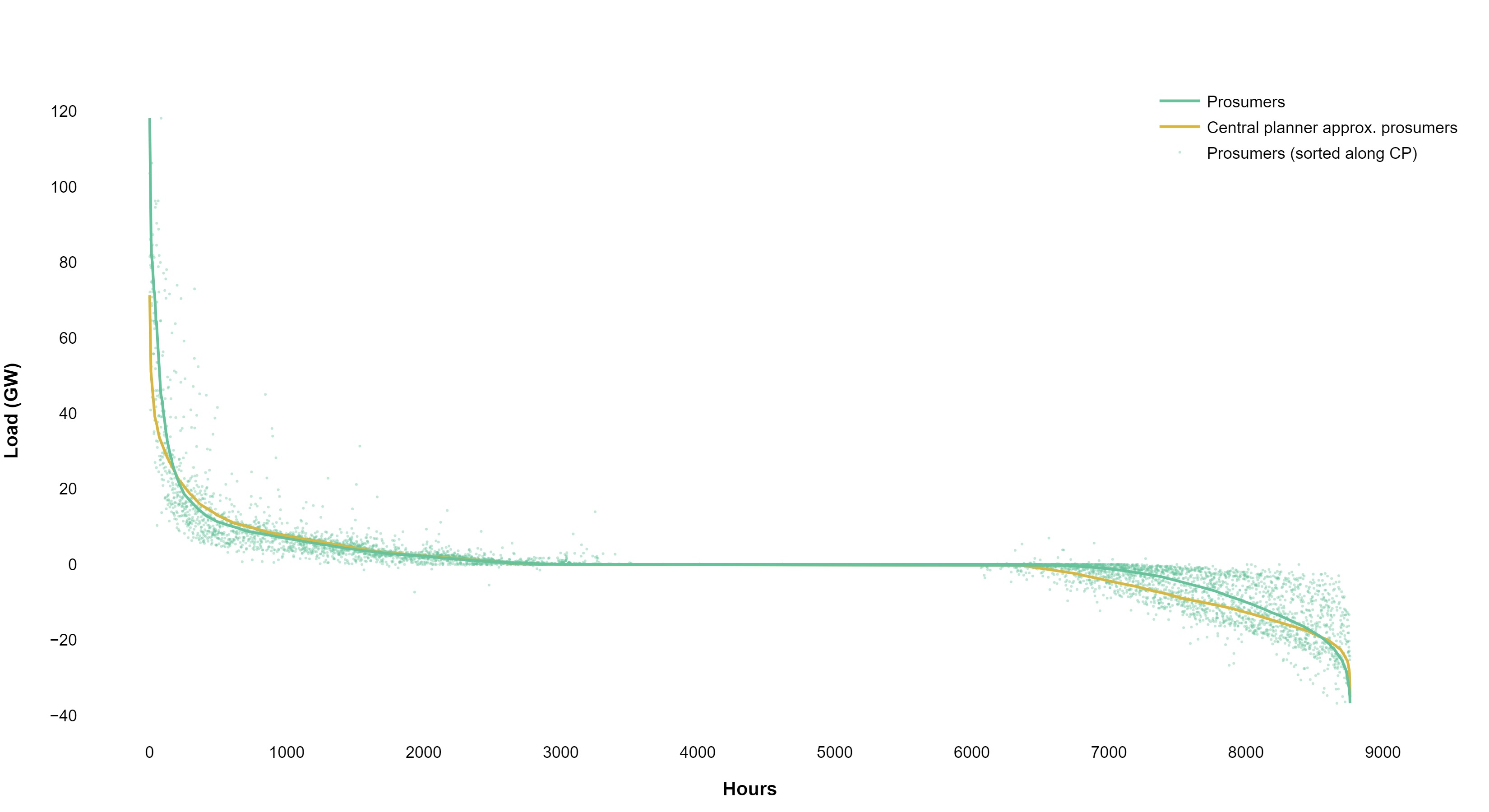}
    \caption{Prosumer net grid consumption (purchased electricity from the grid minus overall self-generated electricity fed into the grid) [\textsc{With BEVs}, real-time pricing with a partly time-varying tariff adder of 20 \euro cts/kWh, 10 million prosumers]}
    \label{fig:dispatch_net_imports_rtp50p}
\end{figure}

\begin{figure}[!ht]
    \centering
    \includegraphics[width=0.95\linewidth]{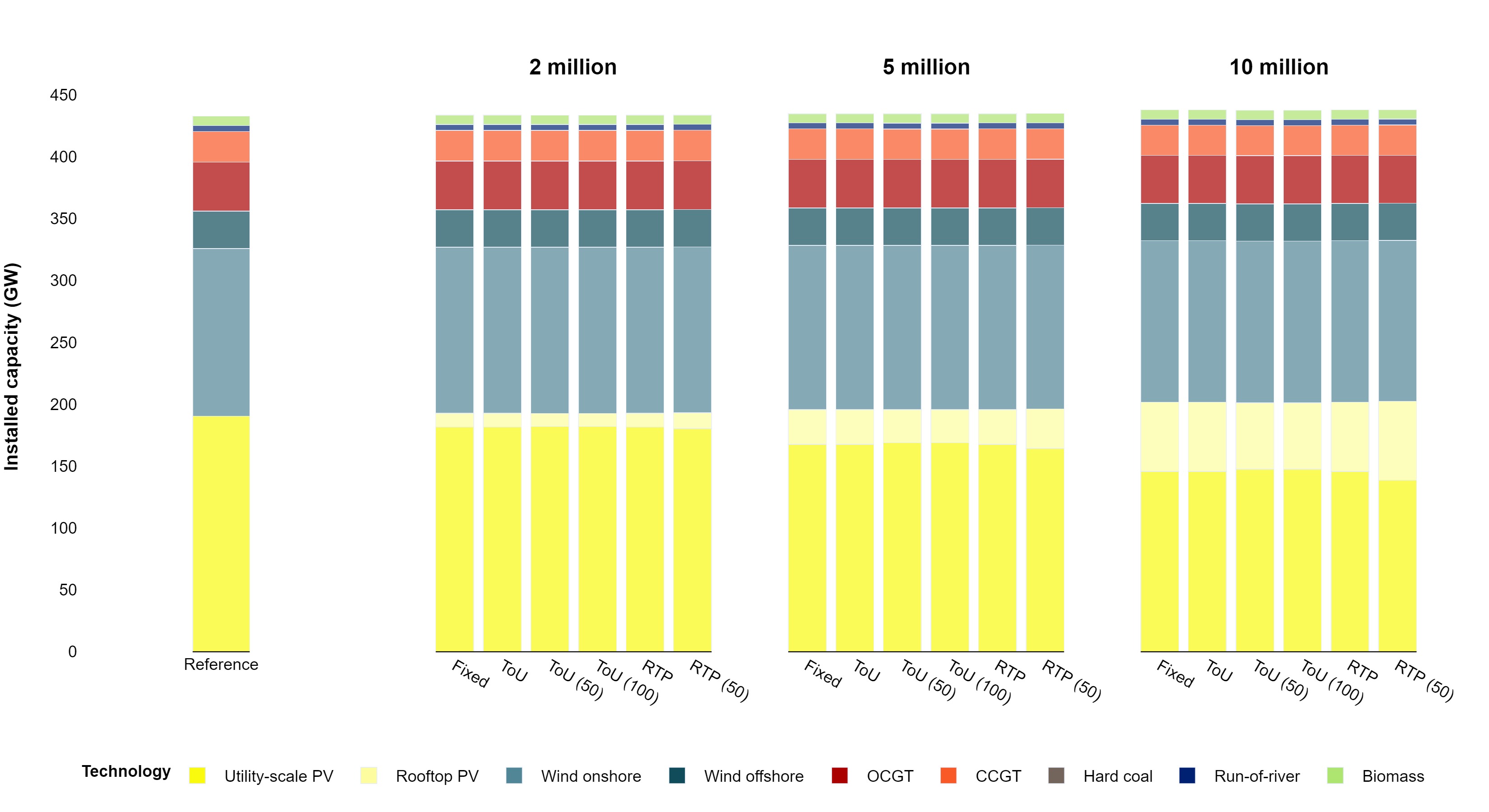}
    \includegraphics[width=0.95\linewidth]{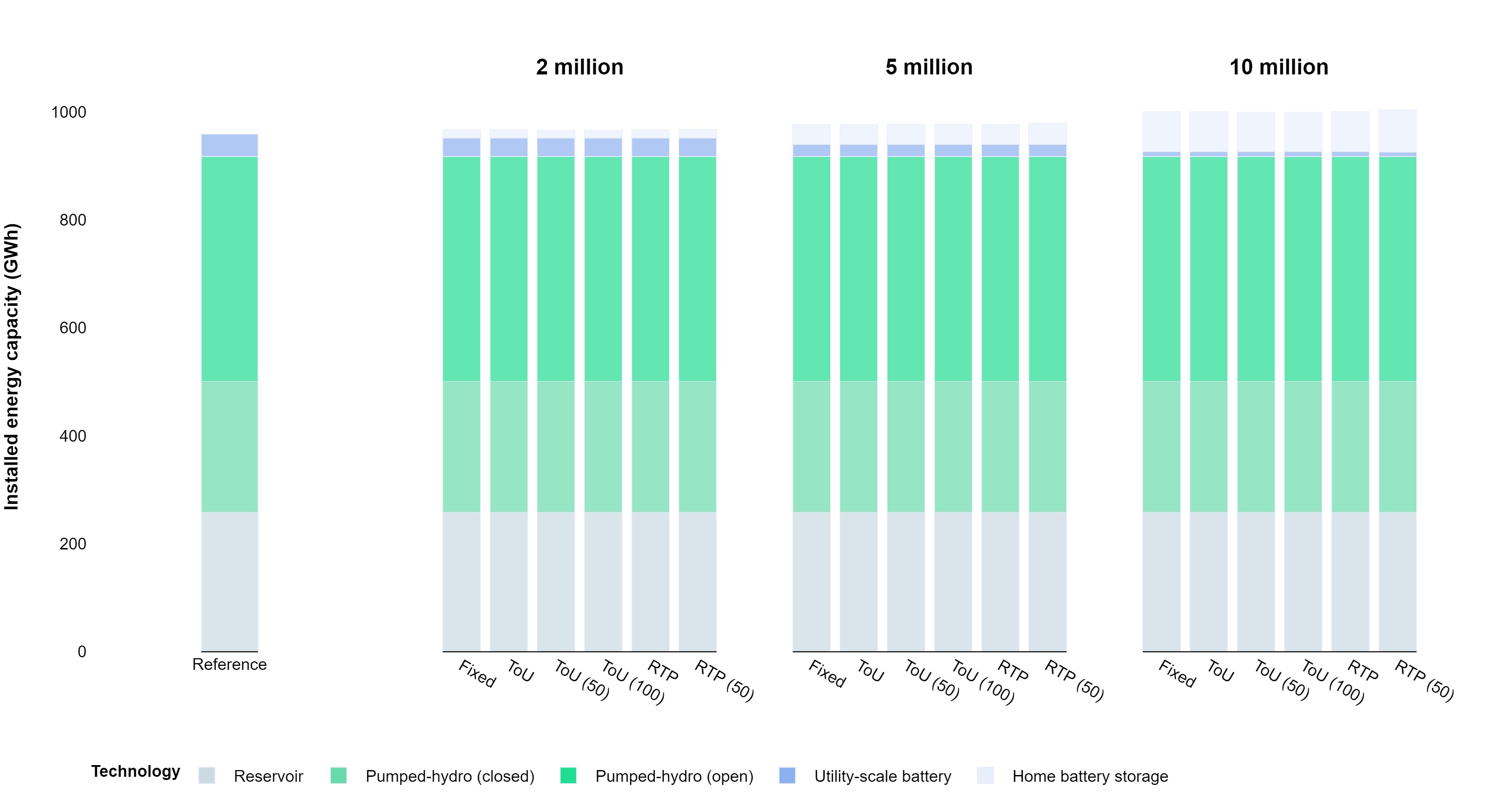}
    \caption{Power generation and storage capacity mix for different numbers of prosumers in the reference and scenarios [\textsc{No BEVs}, tariff adder of 20 euro cts/kWh]}
    \label{fig:generation_storage_mix_noev}
\end{figure}

\begin{figure}[!ht]
    \centering
    \includegraphics[width=0.95\linewidth]{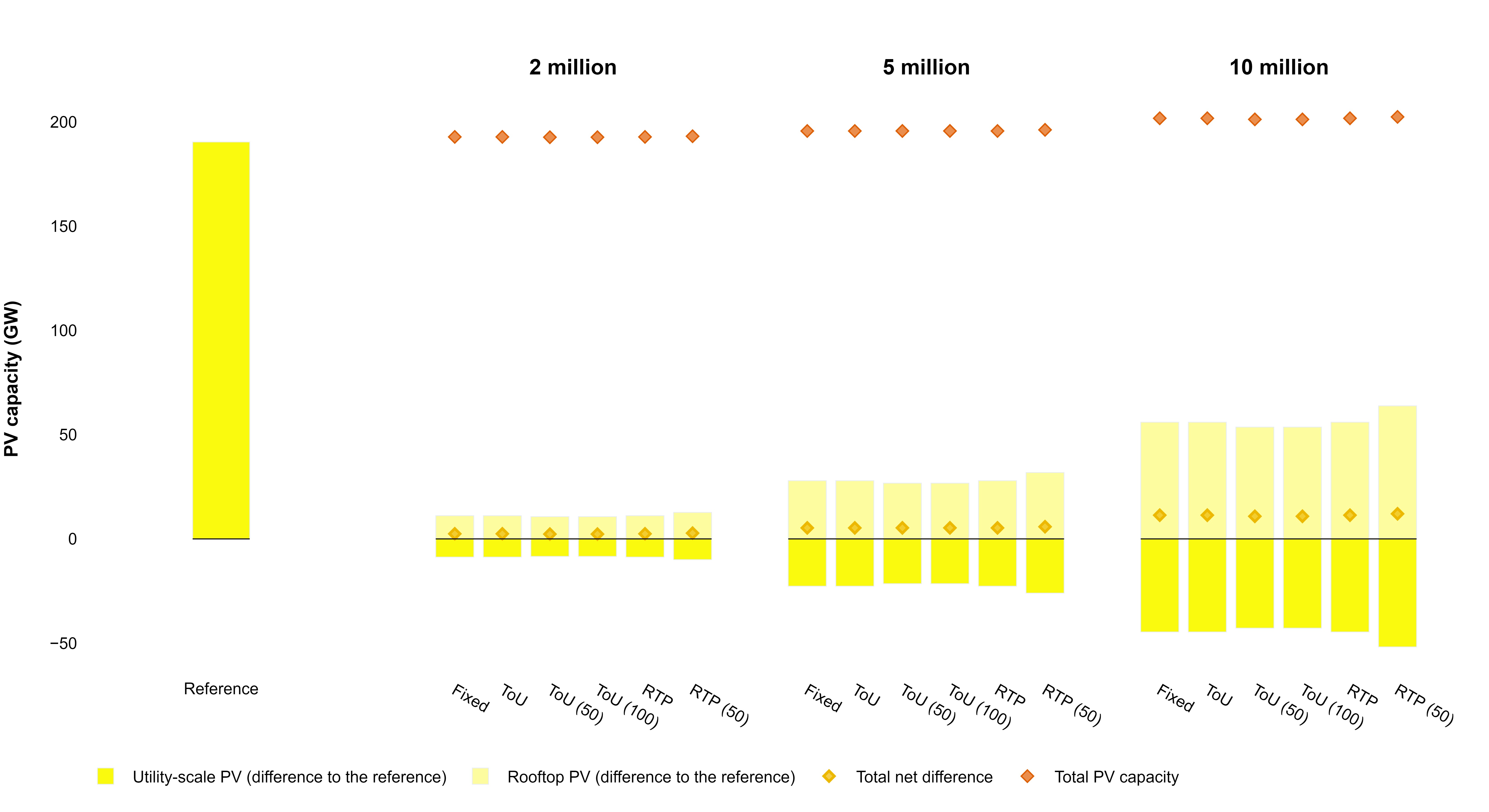}
    \caption{Photovoltaic capacity (in GW) for different numbers of prosumers in the reference and scenarios [\textsc{No BEVs}, tariff adder of 20 \euro cts/kWh]. Note: while the bar for the reference shows an absolute value, all other bars indicate differences to the reference.}
    \label{fig:pv_capacity_noev}
\end{figure}

\begin{figure}[!ht]
    \centering
    \includegraphics[width=0.9\linewidth]{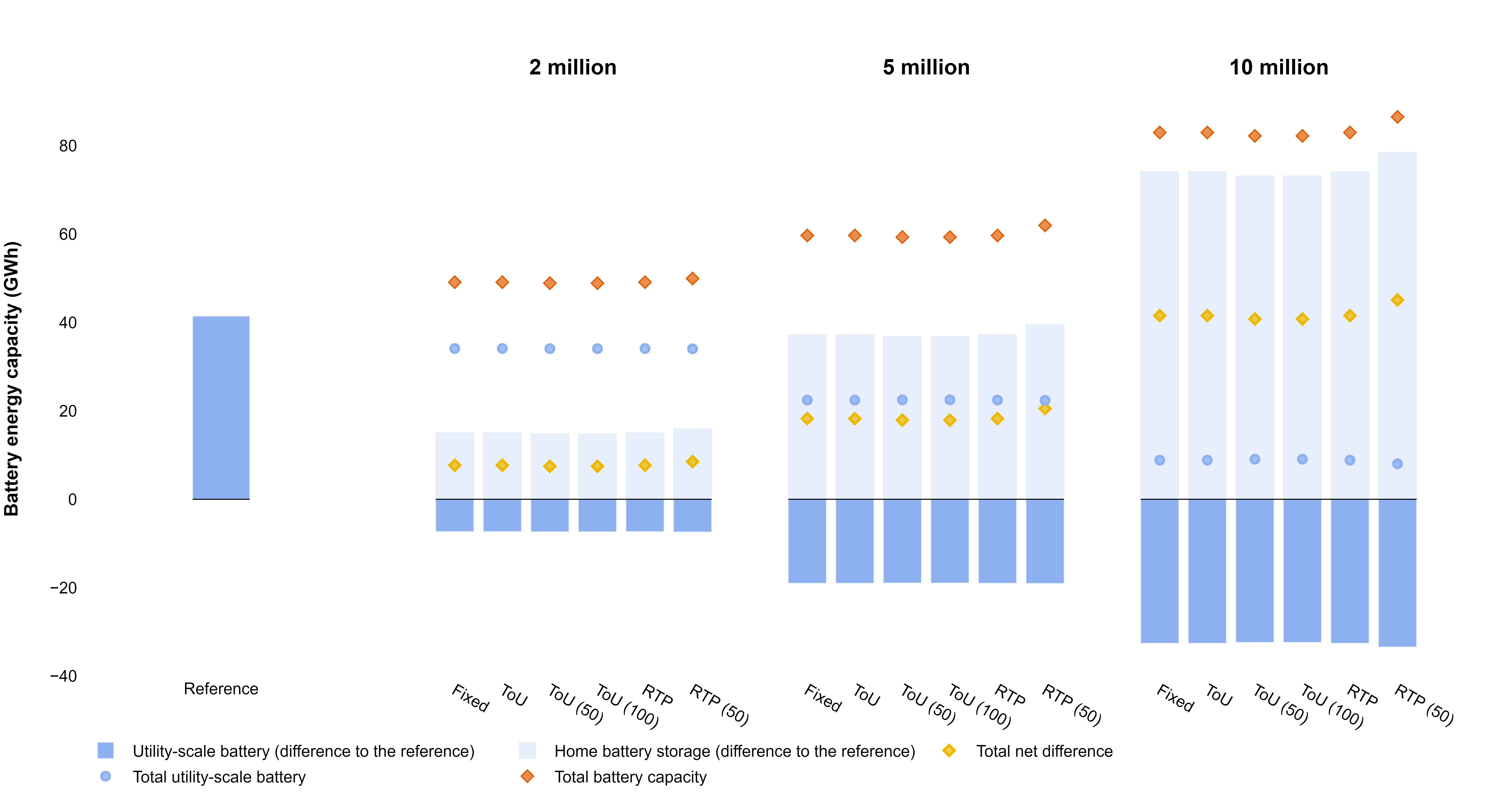}
    \caption{Battery energy capacity (in GWh) for different numbers of prosumers in the reference and scenarios [\textsc{No BEVs}, tariff adder of 20 \euro cts/kWh]. Note: while the bar for the reference shows an absolute value, all other bars indicate differences to the reference.}
    \label{fig:battery_cap_noev}
\end{figure}

\begin{figure}[!ht]
    \centering
    \includegraphics[width=0.95\linewidth]{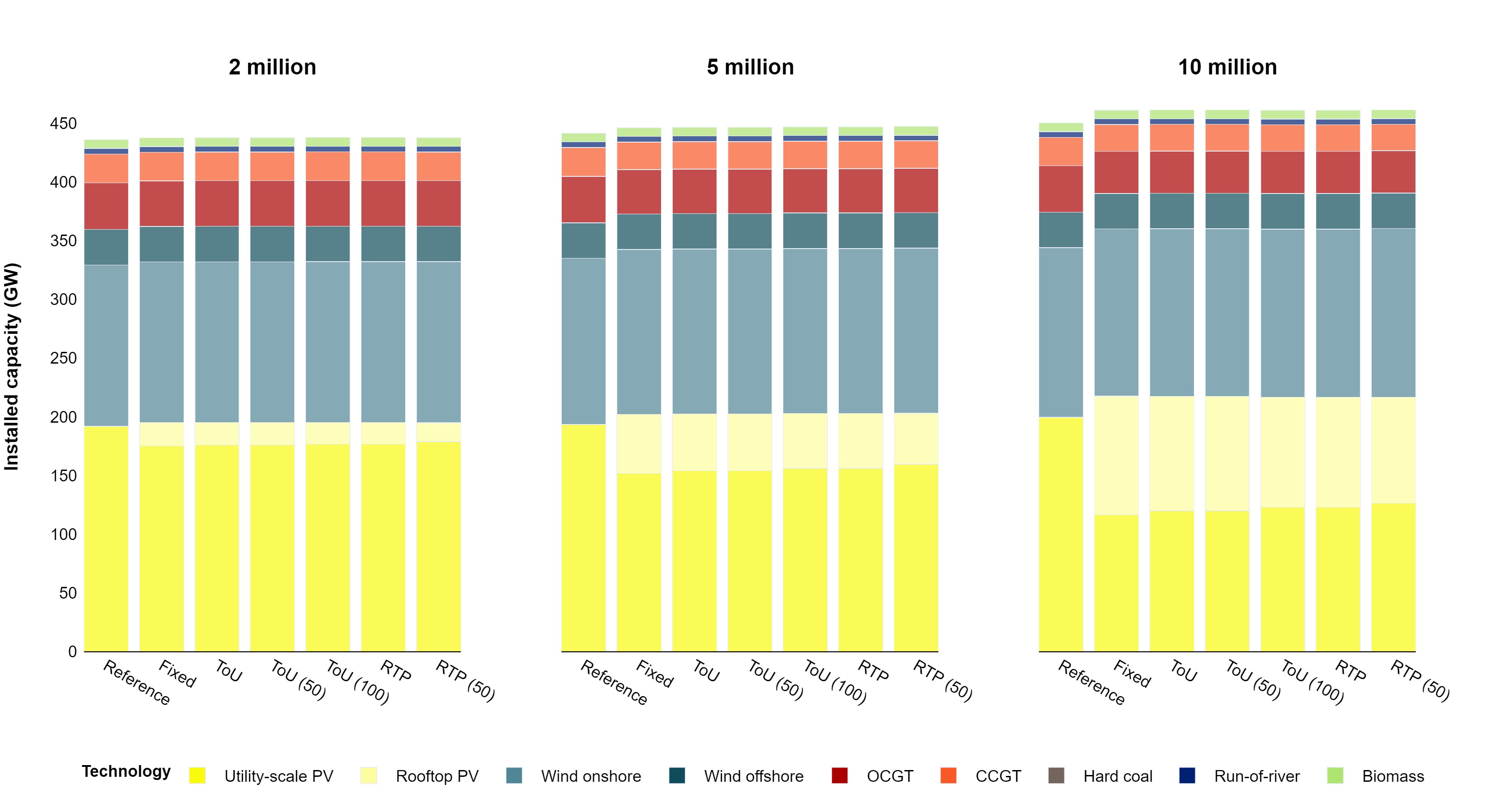}
    \includegraphics[width=0.95\linewidth]{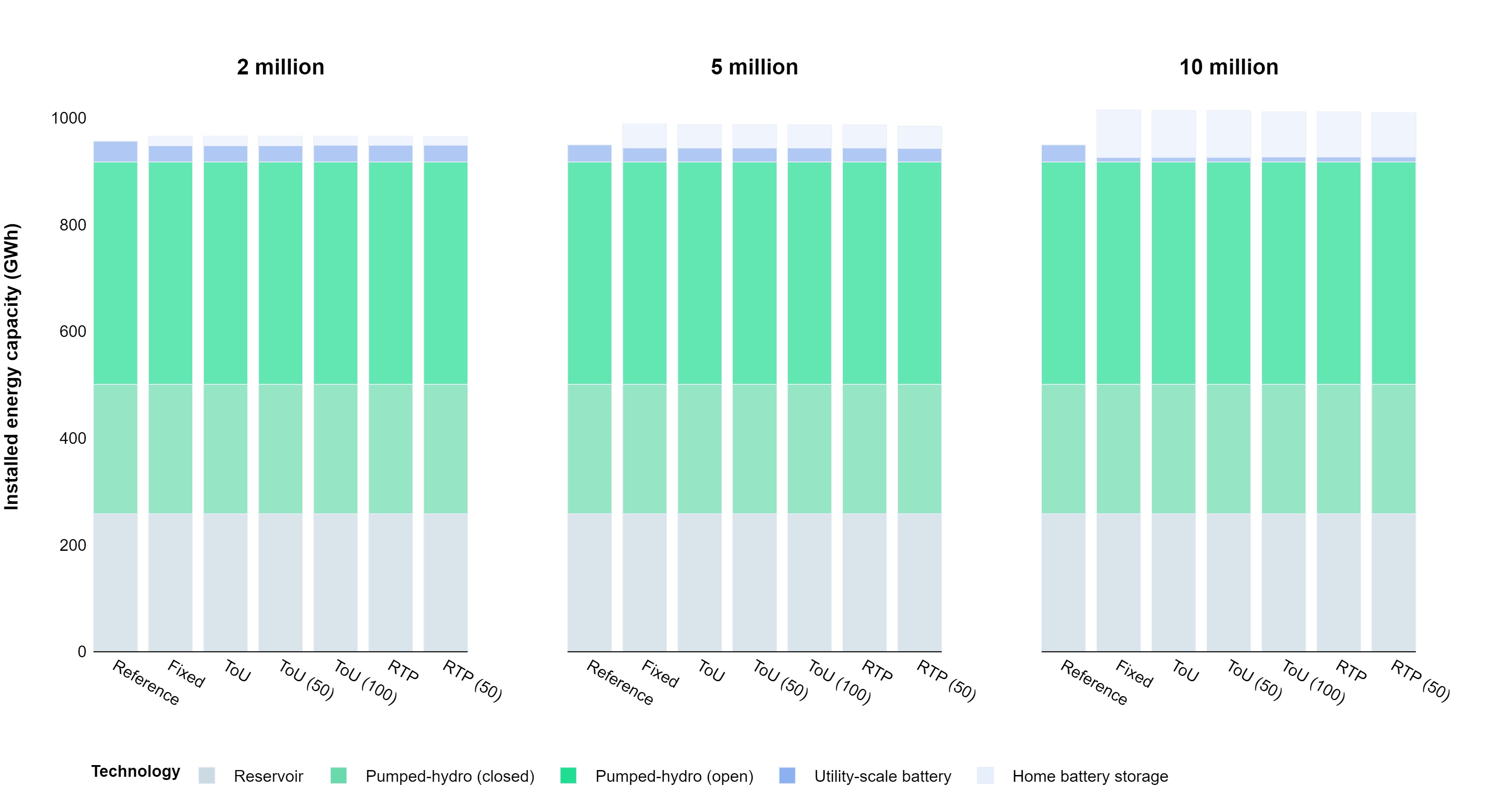}
    \caption{Power generation and storage capacity mix for different numbers of prosumers in the reference and scenarios [\textsc{With BEVs}, tariff adder of 20 euro cts/kWh]}
    \label{fig:generation_storage_mix_withev}
\end{figure}

\begin{figure}[!ht]
    \centering
    \includegraphics[width=0.95\linewidth]{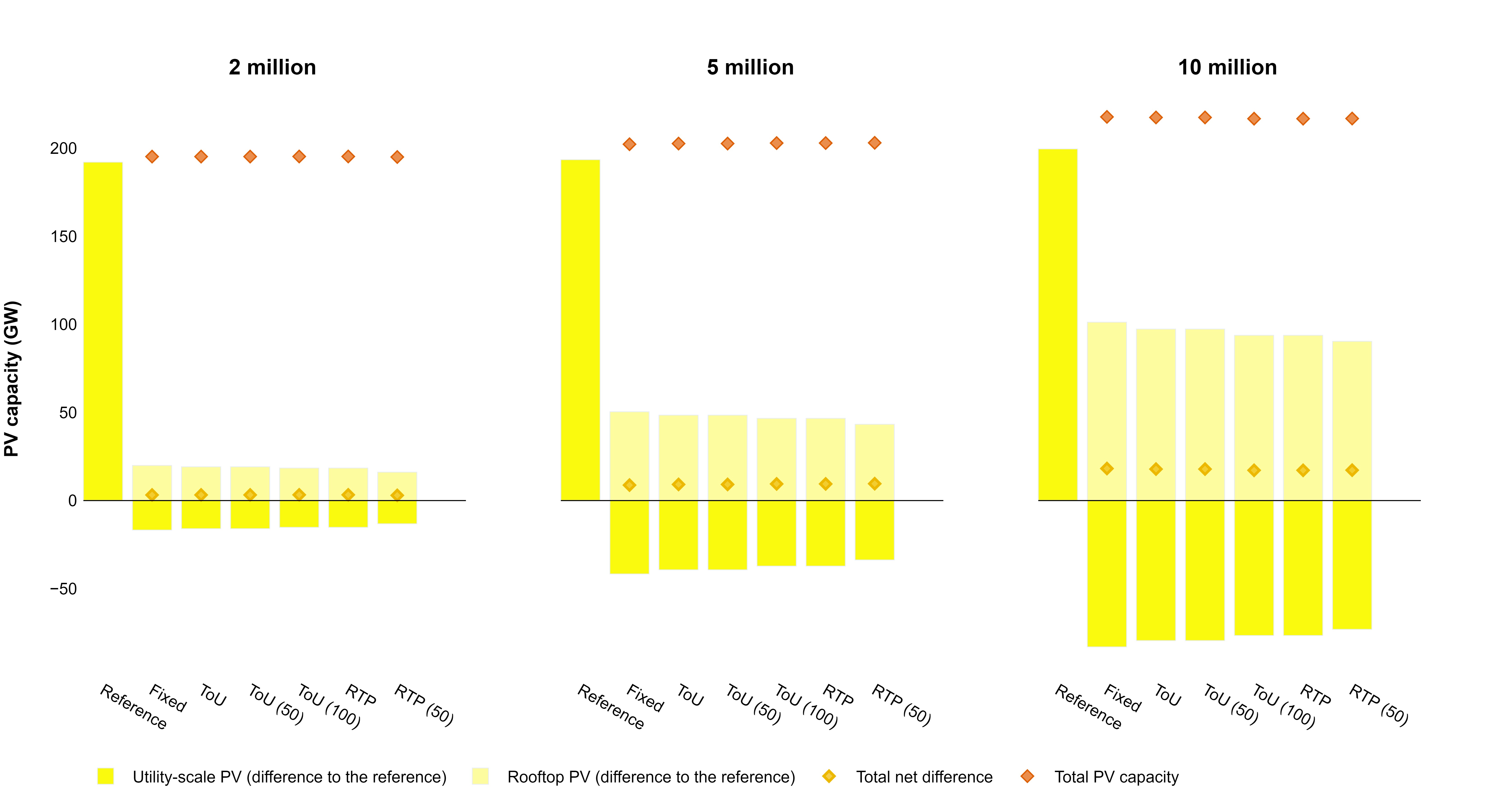}
    \caption{Photovoltaic capacity (in GW) for different numbers of prosumers in the reference and scenarios [\textsc{With BEVs}, tariff adder of 20 euro cts/kWh]. Note: while bars for the reference shows absolute values, all other bars indicate differences to the respective reference.}
    \label{fig:pv_capacity_withev}
\end{figure}

\clearpage
\subsection*{Supplemental Tables}

\begin{table}[!ht]
\centering
\resizebox{\textwidth}{!}{%
\begin{tabular}{@{}
>{\columncolor[HTML]{FFFFFF}}l 
>{\columncolor[HTML]{FFFFFF}}c 
>{\columncolor[HTML]{FFFFFF}}c 
>{\columncolor[HTML]{FFFFFF}}c 
>{\columncolor[HTML]{FFFFFF}}c 
>{\columncolor[HTML]{FFFFFF}}c 
>{\columncolor[HTML]{FFFFFF}}c 
>{\columncolor[HTML]{FFFFFF}}c 
>{\columncolor[HTML]{FFFFFF}}c 
>{\columncolor[HTML]{FFFFFF}}c 
>{\columncolor[HTML]{FFFFFF}}c @{}}
\toprule
 &
  \multicolumn{2}{c}{\cellcolor[HTML]{FFFFFF}\textbf{Capacity bounds}} &
  \cellcolor[HTML]{FFFFFF} &
  \cellcolor[HTML]{FFFFFF} &
  \cellcolor[HTML]{FFFFFF} &
  \cellcolor[HTML]{FFFFFF} &
  \cellcolor[HTML]{FFFFFF} &
  \cellcolor[HTML]{FFFFFF} &
  \cellcolor[HTML]{FFFFFF} &
  \cellcolor[HTML]{FFFFFF} \\ \cmidrule(lr){2-3}
 &
  \textbf{Lower} &
  \textbf{Upper} &
  \multirow{-2}{*}{\cellcolor[HTML]{FFFFFF}\textbf{Interest rates}} &
  \multirow{-2}{*}{\cellcolor[HTML]{FFFFFF}\textbf{Lifetime}} &
  \multirow{-2}{*}{\cellcolor[HTML]{FFFFFF}\textbf{Overnight costs}} &
  \multirow{-2}{*}{\cellcolor[HTML]{FFFFFF}\textbf{Fixed costs}} &
  \multirow{-2}{*}{\cellcolor[HTML]{FFFFFF}\textbf{Variable costs}} &
  \multirow{-2}{*}{\cellcolor[HTML]{FFFFFF}\textbf{Efficiency}} &
  \multirow{-2}{*}{\cellcolor[HTML]{FFFFFF}\textbf{Carbon content}} &
  \multirow{-2}{*}{\cellcolor[HTML]{FFFFFF}\textbf{Fuel costs}} \\
 &
  {[}GW{]} &
  {[}GW{]} &
   &
  {[}years{]} &
  {[}1,000 EUR/MW{]} &
  {[}1,000 EUR/MW{]} &
  {[}EUR/MW{]} &
   &
  {[}t/MWh{]} &
  {[}EUR/MWh{]} \\ \midrule
Hard coal &
  0.00 &
  8.01 &
  0.04 &
  40 &
  4302.034 &
  56.390 &
  3.667 &
  0.405 &
  0.337 &
  9.25 \\
Natural gas (OCGT) &
  8.63 &
  100.00 &
  0.04 &
  25 &
  525.687 &
  9.253 &
  5.618 &
  0.410 &
  0.201 &
  21.06 \\
Natural gas (CCGT) &
  17.27 &
  100.00 &
  0.04 &
  25 &
  991.636 &
  33.214 &
  4.494 &
  0.580 &
  0.201 &
  21.06 \\
Biomass &
  0.00 &
  7.57 &
  0.04 &
  30 &
  2849.799 &
  129.009 &
  0.000 &
  0.468 &
  0.000 &
  13.65 \\
Run-of-river &
  4.73 &
  4.73 &
  0.04 &
  50 &
  3870.256 &
  77.405 &
  0.000 &
  1.000 &
  0.000 &
  0.00 \\
Photovoltaic &
  96.14 &
  +Inf &
  0.04 &
  40 &
  426.945 &
  10.674 &
  0.000 &
  1.000 &
  0.000 &
  0.00 \\
Rooftop photovoltaic &
  0.00 &
  30/75/150 &
  0.04 &
  40 &
  943.774 &
  12.022 &
  0.000 &
  1.000 &
  0.000 &
  0.00 \\
Onshore wind &
  75.37 &
  +Inf &
  0.04 &
  30 &
  1288.302 &
  18.722 &
  2.225 &
  1.000 &
  0.000 &
  0.00 \\
Offshore wind &
  23.63 &
  30.19 &
  0.04 &
  30 &
  2022.373 &
  \cellcolor[HTML]{FFFFFF}43.818 &
  4.494 &
  1.000 &
  0.000 &
  0.00 \\ \bottomrule
\end{tabular}%
}
\caption{Cost and technology parameters for electricity generation technologies}
\label{tab:parameters_gen}
\end{table}

\begin{table}[!h]
\centering
\resizebox{\textwidth}{!}{%
\begin{tabular}{@{}
>{\columncolor[HTML]{FFFFFF}}c 
>{\columncolor[HTML]{FFFFFF}}c 
>{\columncolor[HTML]{FFFFFF}}c 
>{\columncolor[HTML]{FFFFFF}}c 
>{\columncolor[HTML]{FFFFFF}}c 
>{\columncolor[HTML]{FFFFFF}}c 
>{\columncolor[HTML]{FFFFFF}}c 
>{\columncolor[HTML]{FFFFFF}}c 
>{\columncolor[HTML]{FFFFFF}}c 
>{\columncolor[HTML]{FFFFFF}}c 
>{\columncolor[HTML]{FFFFFF}}c 
>{\columncolor[HTML]{FFFFFF}}c 
>{\columncolor[HTML]{FFFFFF}}c 
>{\columncolor[HTML]{FFFFFF}}c 
>{\columncolor[HTML]{FFFFFF}}c @{}}
\toprule
 &
  \multicolumn{4}{c}{\cellcolor[HTML]{FFFFFF}\textbf{Capacity bounds}} &
  \cellcolor[HTML]{FFFFFF} &
  \cellcolor[HTML]{FFFFFF} &
  \cellcolor[HTML]{FFFFFF} &
  \multicolumn{3}{c}{\cellcolor[HTML]{FFFFFF}\textbf{Overnight costs}} &
  \multicolumn{2}{c}{\cellcolor[HTML]{FFFFFF}\textbf{Efficiency}} &
  \multicolumn{2}{c}{\cellcolor[HTML]{FFFFFF}\textbf{Variable costs}} \\ \cmidrule(r){1-5} \cmidrule(l){9-15} 
 &
  \multicolumn{2}{c}{\cellcolor[HTML]{FFFFFF}\textbf{energy}} &
  \multicolumn{2}{c}{\cellcolor[HTML]{FFFFFF}\textbf{power in/out}} &
  \cellcolor[HTML]{FFFFFF} &
  \cellcolor[HTML]{FFFFFF} &
  \cellcolor[HTML]{FFFFFF} &
  \textbf{energy} &
  \textbf{charging power} &
  \textbf{discharging power} &
  \textbf{charging} &
  \textbf{discharging} &
  \textbf{charging} &
  \textbf{discharging} \\
 &
  \textbf{Lower} &
  \textbf{Upper} &
  \textbf{Lower} &
  \textbf{Upper} &
  \multirow{-3}{*}{\cellcolor[HTML]{FFFFFF}\textbf{Interest rates}} &
  \multirow{-3}{*}{\cellcolor[HTML]{FFFFFF}\textbf{Lifetime}} &
  \multirow{-3}{*}{\cellcolor[HTML]{FFFFFF}\textbf{Availability}} &
  \textbf{} &
   &
   &
  \textbf{} &
  \textbf{} &
   &
   \\ \cmidrule(lr){6-8}
 &
  {[}GWh{]} &
  {[}GWh{]} &
  {[}GW{]} &
  {[}GW{]} &
  \textbf{} &
  {[}years{]} &
   &
  {[}1,000 EUR/MWh{]} &
  {[}1,000 EUR/MW{]} &
  {[}1,000 EUR/MW{]} &
   &
   &
  {[}EUR/MW{]} &
  {[}EUR/MW{]} \\ \midrule
\multicolumn{1}{l}{\cellcolor[HTML]{FFFFFF}Lithium-ion batteries} &
  0 &
  +Inf &
  0 &
  +Inf &
  0.04 &
  25 &
  0.99 &
  169.658 &
  191.164 &
  0.011 &
  0.985 &
  0.975 &
  \cellcolor[HTML]{FFFFFF}{\color[HTML]{333333} 1.075} &
  \cellcolor[HTML]{FFFFFF}{\color[HTML]{333333} 1.075} \\
\multicolumn{1}{l}{\cellcolor[HTML]{FFFFFF}Long-duration storage} &
   &
   &
   &
   &
   &
   &
   &
   &
   &
   &
   &
   &
   &
   \\ \cmidrule(r){1-1}
\textit{Electrolysers (PEM)} &
  - &
  - &
  0 &
  +Inf &
  0.04 &
  25 &
  0.97 &
  - &
  730.301 &
  - &
  \cellcolor[HTML]{FFFFFF}{\color[HTML]{333333} 0.585} &
  - &
  0.000 &
  - \\
\textit{Compressor} &
  - &
  - &
  0 &
  +Inf &
  0.04 &
  15 &
  - &
  - &
  95.656 &
  - &
  - &
  - &
  0.000 &
  - \\
\textit{Cavern storage} &
  0 &
  +Inf &
  - &
  - &
  0.04 &
  100 &
  1.00 &
  2.390 &
  - &
  - &
  - &
  - &
  - &
  - \\
\textit{Reconversion} &
  - &
  - &
  0 &
  +Inf &
  0.04 &
  25 &
  0.98 &
  - &
  - &
  538.067 &
  - &
  \cellcolor[HTML]{FFFFFF}{\color[HTML]{333333} 0.41} &
  - &
  5.000 \\
\multicolumn{1}{l}{\cellcolor[HTML]{FFFFFF}Pumped-hydro storage} &
   &
   &
   &
   &
   &
   &
   &
   &
   &
   &
   &
   &
   &
   \\ \cmidrule(r){1-1}
\textit{Closed} &
  242.17 &
  242.17 &
  7.476/7.381 &
  7.476/7.381 &
  0.04 &
  60 &
  1.00 &
  62.952 &
  770.538 &
  770.538 &
  0.89 &
  0.89 &
  0.000 &
  0.000 \\
\textit{Open} &
  416.67 &
  416.67 &
  1.361/1.644 &
  1.361/1.644 &
  0.04 &
  60 &
  1.00 &
  62.952 &
  770.538 &
  770.538 &
  0.89 &
  0.89 &
  0.000 &
  0.000 \\
\multicolumn{1}{l}{\cellcolor[HTML]{FFFFFF}Reservoirs} &
  258.58 &
  258.58 &
  0.000/1.297 &
  0.000/1.297 &
  0.04 &
  60 &
  1.00 &
  62.952 &
  0.000 &
  770.538 &
  1 &
  0.89 &
  0.000 &
  0.000 \\ \bottomrule
\end{tabular}%
}
\caption{Cost and technology parameters for electricity storage technologies}
\label{tab:parameters_sto}
\end{table}

\clearpage
\subsection*{Supplemental Notes}

\textbf{Battery electric vehicles time series} 

\medskip

\textit{emobpy} is an open-source tool used to generate time series for battery electric vehicles at an hourly resolution\autocite{gaete2021dieterpy}. In this paper, we generate twenty profiles that are scaled up to the overall BEV fleet size, i.e. 2, 5 or 10~million BEVs depending on the scenario. We assume an homogeneous fleet of BEVs, taking the Volkswagen~ID.3. model as a benchmark with a 45~kWh battery capacity. Possible trip destinations are work, leisure and home. Charging stations at the workplace and at home are assumed to have a 11~kW power rating, while public chargers are assumed to have a 22~kW power rating. When parking at home or at the workplace, we assume a probability of~1~to find a charger. When driving to leisure destinations, we assume a probability of~0.5 to find a public charger. For further details on the parametrisation, please refer to the GitLab repository.  

\medskip

\textbf{Retail electricity pricing schemes} 

\begin{itemize}
    \item[$\diamond$] \textit{Time-invariant (``Fixed'') tariffs} consist in the sum of the yearly average wholesale electricity price and the tariff adder
    \begin{equation*}
    \forall h \in \mathcal{H}, \quad T_h = \frac{1}{|\mathcal{H}|} \sum_{h} P^{wholesale}_{h} + \text{A} \quad \text{with}~\mathcal{H} = [\![ 1,8760 ]\!] ~\text{and}~\text{A} = \{10,15,20,25\}
    \end{equation*}
    \item[$\diamond$] For \textit{time-of-use tariffs}, we define \textit{ex ante} two periods of time: peak hours (from $8~\text{a.m.}$ to $7~\text{p.m.}$) and off-peak hours (from $8~\text{p.m.}$ to $7~\text{a.m.}$). Within a time period, prices correspond to the sum of the yearly average wholesale price for this period and the tariff adder 
    \begin{align*}
    \forall \mathcal{T} &\in \{\text{peak hours}, \text{off-peak hours}\},~\forall h \in \mathcal{\mathcal{H}_{\mathcal{T}}}, \quad T_{h} = \frac{1}{|\mathcal{H}_{\mathcal{T}}|} \sum_{h \in \mathcal{H}_{\mathcal{T}}} P^{wholesale}_{h} + \text{TA}_{h} \\
    &\text{with} \quad \mathcal{H}_{\mathcal{T}} = \left\{h \in \mathcal{H}= [\![ 1,8760 ]\!]~\rvert~h \in \mathcal{T} \right\} 
    \end{align*}
    \item[$\diamond$] \textit{Real-time tariffs} correspond to the sum of the hourly wholesale price and the tariff adder 
    \begin{equation*}
    \forall h \in \mathcal{H}, \quad T_h = P^{wholesale}_{h} + \text{TA}_{h} \quad \text{with}~\mathcal{H} = [\![ 1,8760 ]\!]
    \end{equation*}
\end{itemize}

For time-of-use and real-time pricing, with consider three possible variants depending on how the tariff adder behaves over time:

\begin{itemize}
    \item[$\circ$] Variant 1 [ToU/RTP]: the tariff adder remains time-invariant;
    \begin{equation*}
    \forall h \in \mathcal{H}, \quad \text{TA}_h = \text{A} \quad \text{with}~\mathcal{H} = [\![ 1,8760 ]\!] ~\text{and}~\text{A} = \{10,15,20,25\}
    \end{equation*}
    \item[$\circ$] Variant 2 [ToU (50)/ RTP (50)]: half of the tariff adder is time-invariant; the other half is time-varying
    \begin{align*}
    \forall h \in \mathcal{H}, \quad \text{TA}_h = \frac{\text{A}}{2} + \frac{P^{wholesale}_{h}}{\bar{P}^{wholesale}} \times \frac{\text{A}}{2} \quad &\text{with}~\mathcal{H} = [\![ 1,8760 ]\!],~\text{A} = \{10,15,20,25\} \\
    &\text{and}~\bar{P}^{wholesale}= \frac{1}{|\mathcal{H}|}\times \sum_{h \in \mathcal{H}} P^{wholesale}_{h}
    \end{align*}
    \item[$\circ$] Variant 3 [ToU (100)/ RTP (100)]: the tariff adder is fully time-varying
    \begin{align*}
    \forall h \in \mathcal{H}, \quad \text{TA}_h = \frac{P^{wholesale}_{h}}{\bar{P}^{wholesale}} \times  \text{A} \quad &\text{with}~\mathcal{H} = [\![ 1,8760 ]\!],~\text{A} = \{10,15,20,25\} \\
    &\text{and}~\bar{P}^{wholesale}= \frac{1}{|\mathcal{H}|}\times \sum_{h \in \mathcal{H}} P^{wholesale}_{h}
    \end{align*}
\end{itemize}

\end{document}